		\documentclass[10pt,journal,compsoc]{IEEEtran}
		\ifCLASSOPTIONcompsoc
		  \usepackage[no compress]{cite}
		\else
		  \usepackage{cite}
		\fi
		\ifCLASSINFOpdf
		\else
		\fi
		\usepackage{mathtools}
		\usepackage{stmaryrd}
		\usepackage{amssymb}
		\usepackage{amsmath}
		\usepackage{moreverb}
		\usepackage{amssymb}
		\usepackage{lineno}
		\usepackage{graphicx}
		\usepackage{amssymb}
		\usepackage{amsmath}
		\usepackage{moreverb}
		\usepackage{amssymb}
		\usepackage{lineno}
		\usepackage{graphicx}
		\graphicspath{ {images/} }
		\usepackage{caption}
		\usepackage{subcaption}
		\usepackage{algpseudocode}
		\usepackage{multirow}
		\usepackage[utf8]{inputenc}
		\usepackage[english]{babel}
		\usepackage{mathtools}
            \usepackage{adjustbox}

        \usepackage{setspace}
        \usepackage{titlesec}
        \titlespacing*{\section}{0pt}{0.5\baselineskip}{0.3\baselineskip}
        \titlespacing*{\subsection}{0pt}{0.4\baselineskip}{0.2\baselineskip}
        \titlespacing*{\subsubsection}{0pt}{0.3\baselineskip}{0.1\baselineskip}
        
		
		\usepackage[colorlinks,bookmarksopen,bookmarksnumbered,citecolor=red,urlcolor=red]{hyperref}
		\usepackage{rotating,booktabs,comment,array,tabu,pdflscape}
		\usepackage{longtable,slashbox,color,colortbl}
		\usepackage{float}
		\usepackage{stmaryrd}
		\usepackage{multirow}
		\usepackage{epstopdf}
		\usepackage{makecell}
		\usepackage{multirow}
		\usepackage{booktabs}
		\usepackage{graphicx}
		\usepackage[table,xcdraw]{xcolor}
		\definecolor{Gray}{gray}{0.9}
		\usepackage[linesnumbered,boxed,lined, ruled]{algorithm2e}
		\let\oldnl\nl
		\newcommand{\nonl}{\renewcommand{\nl}{\let\nl\oldnl}}
		\usepackage{enumitem}
		\usepackage{tabularx}
		\usepackage{ltablex} 
		\usepackage{longtable}
		\usepackage{supertabular}
		\usepackage{color, colortbl}
		\usepackage{csquotes}
		\usepackage{subfloat}
		\usepackage[bottom]{footmisc}
		
		\usepackage{mathrsfs}
		
		\usepackage{stackengine}
		
		\usepackage{amsmath}

\begin{document}
			\fontsize{9.5pt}{11.5pt} 
			%
			\title{Performance and Security Aware Distributed Service Placement in Fog Computing}

	\author{Mohammad Goudarzi, Arash Shaghaghi, Zhiyu Wang, and Rajkumar Buyya
    \IEEEcompsocitemizethanks{
        \IEEEcompsocthanksitem Mohammad Goudarzi is with The Faculty of Information Technology, Monash University, Australia (email: mohammad.goudarzi@monash.edu).
        \IEEEcompsocthanksitem Arash Shaghaghi is with School of Computer Science and Engineering, The University of New South Wales, Australia (email: a.shaghaghi@unsw.edu.au).
        \IEEEcompsocthanksitem Zhiyu Wang and Rajkumar Buyya are with The Quantum Cloud Computing and Distributed Systems (qCLOUDS) Laboratory, School of Computing and Information Systems, The University of Melbourne, Australia (e-mail: zhiywang1@student.unimelb.edu.au, rbuyya@unimelb.edu.au).
    }
}

			\markboth{Submitted to Transactions Journal, VOL.YY, NO.ZZ, YEAR}%
			{Shell \MakeLowercase{\textit{et al.}}: Bare Demo of IEEEtran.cls for Computer Society Journals}
			%

			\IEEEtitleabstractindextext{
\begin{abstract}
The rapid proliferation of Internet of Things (IoT) applications has intensified the demand for efficient and secure service
placement in Fog computing. However, heterogeneous resources, dynamic workloads, and diverse security
requirements make optimal service placement highly challenging. Most existing solutions focus primarily on performance metrics while
overlooking the security implications of deployment decisions. To address these issues, this paper proposes a Security and
Performance-Aware Distributed Deep Reinforcement Learning (SPA-DDRL) framework for joint optimization of service response time
and security compliance in Fog computing. The problem is formulated as a weighted multi-objective optimization task, minimizing
latency while maximizing a security score derived from the security capabilities of Fog nodes. The security score features a new
three-tier hierarchy, where configuration-level checks verify the proper settings, capability-level assessments evaluate the resource
security features, and control-level evaluations enforce stringent policies, thereby guaranteeing compliant solutions while aligning with performance objectives. SPA-DDRL adopts a distributed broker–learner
architecture where multiple brokers perform autonomous service-placement decisions and a centralized learner coordinates global
policy optimization through shared prioritized experiences. SPA-DDRL integrates three key innovations, including Long Short-Term
Memory networks to capture temporal dependencies in dynamic environments, Prioritized Experience Replay to accelerate
convergence by emphasizing critical experiences, and off-policy correction mechanisms to stabilize distributed training. Extensive
experiments based on real IoT workloads demonstrate that SPA-DDRL significantly improves both service response time and
placement security compared with state-of-the-art approaches, where it achieves 16.3\% improvement in response time and converges 33\% faster. Also, SPA-DDRL uniquely maintains consistent, feasible
security-compliant solutions across all system scales, while baseline techniques fail or show performance degradation.

\end{abstract}

	}

\maketitle
 
\IEEEdisplaynontitleabstractindextext

%
\IEEEpeerreviewmaketitle

\IEEEraisesectionheading{\section{Introduction}\label{sec:introduction}}
\IEEEPARstart{I}{nternet} of Things (IoT) devices have become ubiquitous, spanning diverse domains such as smart healthcare, cities, and intelligent transportation \cite{he2025integrating}. These applications aim to deliver efficient solutions by analyzing diverse data streams. Consequently, these computationally intensive services impose substantial demands on computing and communication resources to ensure operational integrity \cite{goudarzi2021distributedDDRL}. However, resource-constrained IoT devices cannot process massive data streams within strict latency bounds. Thus, they offload tasks to surrogate infrastructure, a paradigm known as service placement \cite{du2025online}.
\par
While Cloud computing provides elastic resources for IoT deployment \cite{deng2021fogbus2}, its inherent latency and bandwidth constraints render it unsuitable for time-sensitive applications. Fog computing addresses this deficiency by deploying Fog Servers (FSs) at the network edge, forming a hierarchical architecture that effectively manages both computation-intensive and latency-critical workloads \cite{goudarzi2019fog,al2022ai}.
\par
However, service placement in heterogeneous Fog environments confronts dual challenges. First, the limited capacity of FSs relative to Cloud Servers creates resource contention, necessitating strategic placement to balance performance under heavy loads \cite{al2025real}. Second, Fog's open architecture exposes the system to severe security vulnerabilities (e.g., Denial-of-Service (DoS) attacks), with high-performance nodes often serving as prime targets \cite{kong2022edge, deng2025secure, wang2022design}. Since conventional strategies typically prioritize performance over security, leaving critical infrastructure exposed \cite{sun2022security}, there is a fundamental need for approaches that jointly optimize performance efficiency and security compliance.
\par
Given the inherent stochasticity of Fog environments, Deep Reinforcement Learning (DRL) emerges as a robust paradigm for adaptive decision-making, enabling the derivation of optimal policies without a priori system knowledge \cite{huang2025energy}. However, standard DRL implementations face significant scalability and efficiency hurdles. Centralized approaches often incur prohibitive convergence times and high exploration overheads in heterogeneous, high-dimensional environments. Furthermore, existing distributed DRL techniques frequently exhibit poor data utilization, failing to effectively leverage experience trajectories across dispersed actors to achieve global optimality.
\par
To overcome these limitations, we propose the Security and Performance Aware Distributed Deep Reinforcement Learning (SPA-DDRL) framework, jointly optimizing security compliance and service latency. It adopts a distributed broker-learner architecture where autonomous brokers execute decentralized placement decisions, while a centralized learner orchestrates global policy updates. The framework synergizes three core mechanisms: Long Short-Term Memory (LSTM) networks to capture temporal dependencies, Prioritized Experience Replay (PER) to accelerate learning from critical transitions, and off-policy correction to mitigate policy divergence among distributed actors. This design effectively reconciles decision-making scalability with learning stability in heterogeneous Fog environments.

\begin{itemize}
    \item We propose a novel service placement approach framed as a weighted optimization problem. This approach strives to achieve two key objectives: minimizing the response time of services and maximizing the overall security score of the placement. The security score is determined by rigorously evaluating the security capabilities of the available Fog resources.
    \item We put forward SPA-DDRL, a distributed DRL technique for dynamic and stochastic Fog environments. Built upon the Actor-Critic architecture, SPA-DDRL efficiently leverages experience data collected from multiple distributed brokers to train a superior service placement model. Moreover, we design a reward function for SPA-DDRL to jointly optimize the response time of services, represented as Directed Acyclic Graphs (DAGs), and the service placement security score. 
    \item To enhance learning performance, we integrate LSTM networks into Actor-Critic architecture for capturing temporal dependencies in dynamic Fog environments, employ PER to focus learning on critical experiences based on Temporal Difference (TD) errors, and implement off-policy correction mechanisms (e.g., importance sampling and gradient clipping) to address policy divergence between distributed brokers and learner.
    \item To thoroughly evaluate SPA-DDRL, we conduct extensive experiments using a diverse set of synthetic DAGs, derived from real-world IoT services, and compare its performance against related techniques.
\end{itemize}

\par
The remainder of this paper is organized as follows. Section~\ref{relatedw} reviews related literature. Section~\ref{system} presents the system model and the three-tier security hierarchy, while Section~\ref{sec:DRLModel} formalizes the service placement problem as a Markov Decision Process (MDP). Section~\ref{sec:placement} details the proposed SPA-DDRL framework while Section~\ref{sec:evaluation} provides a comprehensive experimental evaluation against existing techniques. Section~\ref{conclusion} concludes the paper with directions for future work.
\section{Related Work}
\label{relatedw}
Several studies (e.g., \cite{sun2022security,mann2020secure,elgendy2019resource,casola2020security,javanmardi2023s,mann2021security,wang2022design,singh2019scheduling,zhang2024security,rahmani2024novel,ebrahim2023privacy,sun2024secure,mohammadi2025security,du2025secure,thangaraj2024msco}) have delved into optimizing service placement for IoT services in Fog computing, primarily focusing on performance metrics such as response time and energy efficiency. However, most of these studies overlook the crucial aspect of security in service placement within Edge and Fog computing environments. This section addresses this critical but often neglected dimension by exploring service placement strategies that prioritize both the performance and security of IoT services.
\par
The existing literature on security-aware service placement can be broadly categorized into conventional optimization approaches and learning-based approaches. Regarding conventional optimization, Sun et al.~\cite{sun2022security} proposed a heuristic for IoT task scheduling that ranks tasks based on a probabilistic risk model to satisfy security requirements and strict deadlines. Mann et al.~\cite{mann2020secure} modeled service placement and security control selection as a constraint satisfaction problem, utilizing Gurobi to optimize hardware and software-level security configurations. Similarly, Elgendy et al.~\cite{elgendy2019resource} applied a branch-and-bound technique to optimize computation offloading with integrated AES encryption, jointly minimizing latency and energy consumption. Casola et al.~\cite{casola2020security} developed a greedy strategy for Industrial IoT placement, balancing cost and performance while enforcing specific security policy constraints. Javanmardi et al.~\cite{javanmardi2023s} introduced an SDN-based metaheuristic scheduler that mitigates DDoS risks by dynamically monitoring device security status and isolating suspicious nodes. Extending their previous work, Mann et al.~\cite{mann2021security} addressed joint application placement and user assignment under module- and user-level location security constraints using exact solvers. Wang et al.~\cite{wang2022design} leveraged metaheuristics to concurrently optimize response time and security risks for distributed data placement, operating under the premise that high-performance nodes face elevated attack probabilities. Singh et al.~\cite{singh2019scheduling} designed a tag-based heuristic that maps services to resources (e.g., trusted private clouds) based on security classifications while adhering to execution deadlines.
\par
Transitioning to learning-based and dynamic approaches, Zhang et al.~\cite{zhang2024security} developed an action-constrained Deep Q-Network (DQN) for multi-cloud edge networks, integrating AES and RSA encryption constraints directly into the computation offloading process. Rahmani et al.~\cite{rahmani2024novel} proposed a hybrid framework combining Asynchronous Advantage Actor-Critic (A3C) with Analytic Hierarchy Process (AHP) for blockchain-enabled MEC, balancing latency, energy, and secure transaction validation. Ebrahim et al.~\cite{ebrahim2023privacy} addressed privacy-preserving load balancing using a Double DQN (DDQN) framework that manages workload distribution under partial observability without exposing sensitive node details. Sun et al.~\cite{sun2024secure} formulated resource allocation as a Markov decision process, utilizing an action-constrained DQN that dynamically adjusts protection levels (e.g., RSA, MD5) based on task sensitivity. Mohammadi et al.~\cite{mohammadi2025security} enhanced the NSGA-II algorithm with sigma scaling to solve multi-objective resource allocation in device-to-device Fog systems, optimizing delay, energy, and breach costs. Du et al.~\cite{du2025secure} designed an enhanced Genetic Algorithm (GA) for Ocean IoT, incorporating a correction operator and security quantification model to ensure feasible, secure offloading. Finally, Thangaraj et al.~\cite{thangaraj2024msco} presented a hybrid GA-PSO mechanism for blockchain-enabled Fog, selecting authorized servers based on mobility patterns to ensure data integrity and minimize latency.
%
\begin{table*}[]
\caption{A qualitative comparison of related works with ours}
\label{tab:relatedwork}
\resizebox{\textwidth}{!}{%
\begin{tabular}{|cccccccccccccccccccc|}
\hline
\multicolumn{1}{|c|}{} & \multicolumn{4}{c|}{Service Properties} & \multicolumn{4}{c|}{Environmental Properties} & \multicolumn{3}{c|}{Problem Formulation Properties} & \multicolumn{5}{c|}{Decision Engine Properties} & \multicolumn{2}{c|}{Security Properties} & \multicolumn{1}{c|}{Evaluation Properties} \\ \cline{2-20} 
\multicolumn{1}{|c|}{\multirow{-2}{*}{Properties}} & \multicolumn{1}{c|}{Structure} & \multicolumn{1}{c|}{Number} & \multicolumn{1}{c|}{Constraints} & \multicolumn{1}{c|}{Heter} & \multicolumn{1}{c|}{User/IoT} & \multicolumn{1}{c|}{Hierarchy} & \multicolumn{1}{c|}{Edge Num} & \multicolumn{1}{c|}{Heter} & \multicolumn{1}{c|}{Formula} & \multicolumn{1}{c|}{Parameters} & \multicolumn{1}{c|}{Constraints} & \multicolumn{1}{c|}{\begin{tabular}[c]{@{}c@{}}Placement\\ Layer\end{tabular}} & \multicolumn{1}{c|}{Solver} & \multicolumn{1}{c|}{CMP} & \multicolumn{1}{c|}{ADAP} & \multicolumn{1}{c|}{SCAL} & \multicolumn{1}{c|}{Main Goal} & \multicolumn{1}{c|}{Mitigation} & Evaluation Metric \\ \hline
\multicolumn{1}{|c|}{\cite{sun2022security}} & \multicolumn{1}{c|}{Service} & \multicolumn{1}{c|}{Multiple} & \multicolumn{1}{c|}{Dependent} & \multicolumn{1}{c|}{\checkmark} & \multicolumn{1}{c|}{Multiple} & \multicolumn{1}{c|}{\begin{tabular}[c]{@{}c@{}}Three\\ Layer\end{tabular}} & \multicolumn{1}{c|}{Multiple} & \multicolumn{1}{c|}{\checkmark} & \multicolumn{1}{c|}{MINLP} & \multicolumn{1}{c|}{\begin{tabular}[c]{@{}c@{}}Risk\\probability\end{tabular}} & \multicolumn{1}{c|}{Deadline} & \multicolumn{1}{c|}{Edge} & \multicolumn{1}{c|}{Heuristic} & \multicolumn{1}{c|}{Low} & \multicolumn{1}{c|}{Low} & \multicolumn{1}{c|}{Mid} & \multicolumn{1}{c|}{\begin{tabular}[c]{@{}c@{}}Meeting security requirements \\ of different applications.\end{tabular}} & \multicolumn{1}{c|}{\begin{tabular}[c]{@{}c@{}}Ranking tasks based on\\security requirements and\\ their placement on  best computing nodes \end{tabular}} & \begin{tabular}[c]{@{}c@{}}Response Time,\\ Risk Level,\\ Security Level\end{tabular} \\ \hline
\multicolumn{1}{|c|}{\cite{mann2020secure}} & \multicolumn{1}{c|}{Service} & \multicolumn{1}{c|}{Multiple} & \multicolumn{1}{c|}{Dependent} & \multicolumn{1}{c|}{$\times$} & \multicolumn{1}{c|}{Single} & \multicolumn{1}{c|}{\begin{tabular}[c]{@{}c@{}}Two\\ Layer\end{tabular}} & \multicolumn{1}{c|}{Multiple} & \multicolumn{1}{c|}{\checkmark} & \multicolumn{1}{c|}{MIQP} & \multicolumn{1}{c|}{Performance} & \multicolumn{1}{c|}{\begin{tabular}[c]{@{}c@{}}CPU\\ Req\end{tabular}} & \multicolumn{1}{c|}{NA} & \multicolumn{1}{c|}{\begin{tabular}[c]{@{}c@{}}\begin{tabular}[c]{@{}c@{}}Optimization\\Solver\end{tabular}\end{tabular}} & \multicolumn{1}{c|}{High} & \multicolumn{1}{c|}{Low} & \multicolumn{1}{c|}{Low} & \multicolumn{1}{c|}{\begin{tabular}[c]{@{}c@{}}application task placement and\\configuration of security controls.\end{tabular}} & \multicolumn{1}{c|}{\begin{tabular}[c]{@{}c@{}}Categorizing application and servers' security controls\\as a CSP problem, and finding a satisfactory solution.\end{tabular}} & Response Time \\ \hline
\multicolumn{1}{|c|}{\cite{elgendy2019resource}} & \multicolumn{1}{c|}{Task} & \multicolumn{1}{c|}{Multiple} & \multicolumn{1}{c|}{Independent} & \multicolumn{1}{c|}{\checkmark} & \multicolumn{1}{c|}{Multiple} & \multicolumn{1}{c|}{\begin{tabular}[c]{@{}c@{}}Two\\ Layer\end{tabular}} & \multicolumn{1}{c|}{Single} & \multicolumn{1}{c|}{$\times$} & \multicolumn{1}{c|}{ILP} & \multicolumn{1}{c|}{\begin{tabular}[c]{@{}c@{}}Time,\\ Energy\end{tabular}} & \multicolumn{1}{c|}{\begin{tabular}[c]{@{}c@{}}Time,\\ Energy\end{tabular}} & \multicolumn{1}{c|}{Edge} & \multicolumn{1}{c|}{Branch\&B} & \multicolumn{1}{c|}{High} & \multicolumn{1}{c|}{Low} & \multicolumn{1}{c|}{Low} & \multicolumn{1}{c|}{\begin{tabular}[c]{@{}c@{}}Protecting sensitive placement\\ information.\end{tabular}} & \multicolumn{1}{c|}{\begin{tabular}[c]{@{}c@{}}AES encryption technique to protect\\ sensitive information.\end{tabular}} & \begin{tabular}[c]{@{}c@{}}Response Time,\\ Energy\end{tabular} \\ \hline
\multicolumn{1}{|c|}{\cite{casola2020security}} & \multicolumn{1}{c|}{Service} & \multicolumn{1}{c|}{Multiple} & \multicolumn{1}{c|}{Dependent} & \multicolumn{1}{c|}{NA} & \multicolumn{1}{c|}{Single} & \multicolumn{1}{c|}{\begin{tabular}[c]{@{}c@{}}Three\\ Layer\end{tabular}} & \multicolumn{1}{c|}{Multiple} & \multicolumn{1}{c|}{\checkmark} & \multicolumn{1}{c|}{INLP} & \multicolumn{1}{c|}{Cost, Security} & \multicolumn{1}{c|}{Security} & \multicolumn{1}{c|}{Edge} & \multicolumn{1}{c|}{Greedy} & \multicolumn{1}{c|}{Low-Mid} & \multicolumn{1}{c|}{Low} & \multicolumn{1}{c|}{Mid} & \multicolumn{1}{c|}{\begin{tabular}[c]{@{}c@{}}Satisfying the security requirements\\  of tasks.\end{tabular}} & \multicolumn{1}{c|}{\begin{tabular}[c]{@{}c@{}} Defining security controls for each task and\\ satisfying them based on offered security levels.\end{tabular}} & Cost \\ \hline
\multicolumn{1}{|c|}{\cite{javanmardi2023s}} & \multicolumn{1}{c|}{NA} & \multicolumn{1}{c|}{Multiple} & \multicolumn{1}{c|}{Independent} & \multicolumn{1}{c|}{NA} & \multicolumn{1}{c|}{Multiple} & \multicolumn{1}{c|}{\begin{tabular}[c]{@{}c@{}}Three\\ Layer\end{tabular}} & \multicolumn{1}{c|}{Multiple} & \multicolumn{1}{c|}{\checkmark} & \multicolumn{1}{c|}{NA} & \multicolumn{1}{c|}{\begin{tabular}[c]{@{}c@{}}Load Balance,\\ Delay\end{tabular}} & \multicolumn{1}{c|}{$\times$} & \multicolumn{1}{c|}{Edge} & \multicolumn{1}{c|}{\begin{tabular}[c]{@{}c@{}}Metaheuristic\\ (NSGA3)\end{tabular}} & \multicolumn{1}{c|}{Mid} & \multicolumn{1}{c|}{Low} & \multicolumn{1}{c|}{Mid} & \multicolumn{1}{c|}{\begin{tabular}[c]{@{}c@{}}Resource management framework to\\consider security status of IoT devices\\and exclude suspicious ones.\end{tabular}} & \multicolumn{1}{c|}{\begin{tabular}[c]{@{}c@{}}Employing outcomes of IDS methods as the input\\of fuzzy function to put away the IoT devices\\lunching DDoS and scanning attacks.\end{tabular}} & \begin{tabular}[c]{@{}c@{}}Response Time,\\ Network Usage\end{tabular} \\ \hline
\multicolumn{1}{|c|}{\cite{mann2021security}} & \multicolumn{1}{c|}{Service} & \multicolumn{1}{c|}{Multiple} & \multicolumn{1}{c|}{Independent} & \multicolumn{1}{c|}{\checkmark} & \multicolumn{1}{c|}{Multiple} & \multicolumn{1}{c|}{\begin{tabular}[c]{@{}c@{}}Three\\ Layer\end{tabular}} & \multicolumn{1}{c|}{Single} & \multicolumn{1}{c|}{$\times$} & \multicolumn{1}{c|}{QCMIP} & \multicolumn{1}{c|}{Delay} & \multicolumn{1}{c|}{\begin{tabular}[c]{@{}c@{}}Security,\\ \\ Capacity\end{tabular}} & \multicolumn{1}{c|}{NA} & \multicolumn{1}{c|}{\begin{tabular}[c]{@{}c@{}}Optimization\\Solver\end{tabular}} & \multicolumn{1}{c|}{High} & \multicolumn{1}{c|}{Low} & \multicolumn{1}{c|}{Low} & \multicolumn{1}{c|}{\begin{tabular}[c]{@{}c@{}}Application placement and user\\ assignment while considering different\\ security and privacy constraints\end{tabular}} & \multicolumn{1}{c|}{\begin{tabular}[c]{@{}c@{}}Considering module-level location, user-level location,\\co-location, and k-anonymity constraints.\end{tabular}} & Response Time \\ \hline
\multicolumn{1}{|c|}{\cite{wang2022design}} & \multicolumn{1}{c|}{\begin{tabular}[c]{@{}c@{}}Service\\ (Data)\end{tabular}} & \multicolumn{1}{c|}{Multiple} & \multicolumn{1}{c|}{Independent} & \multicolumn{1}{c|}{\checkmark} & \multicolumn{1}{c|}{Multiple} & \multicolumn{1}{c|}{\begin{tabular}[c]{@{}c@{}}Three\\ Layer\end{tabular}} & \multicolumn{1}{c|}{Multiple} & \multicolumn{1}{c|}{\checkmark} & \multicolumn{1}{c|}{NA} & \multicolumn{1}{c|}{\begin{tabular}[c]{@{}c@{}}Time,\\ Security\end{tabular}} & \multicolumn{1}{c|}{$\times$} & \multicolumn{1}{c|}{NA} & \multicolumn{1}{c|}{\begin{tabular}[c]{@{}c@{}}Metaheuristic\\ (Evolutionary\\Algorithm)\end{tabular}} & \multicolumn{1}{c|}{Mid} & \multicolumn{1}{c|}{Mid} & \multicolumn{1}{c|}{Mid} & \multicolumn{1}{c|}{\begin{tabular}[c]{@{}c@{}}Minimizing the response time of\\read/write of distributed files on FSs, while\\ minimizing the risk of attacks.\end{tabular}} & \multicolumn{1}{c|}{\begin{tabular}[c]{@{}c@{}}Tradeoff between response time and security by\\ defining a security risk score based on\\ the centrality of the FSs.\end{tabular}} & \begin{tabular}[c]{@{}c@{}}Response Time,\\ Successful\\  Attacks\end{tabular} \\ \hline
\multicolumn{1}{|c|}{\cite{singh2019scheduling}} & \multicolumn{1}{c|}{Service} & \multicolumn{1}{c|}{Multiple} & \multicolumn{1}{c|}{Independent} & \multicolumn{1}{c|}{\checkmark} & \multicolumn{1}{c|}{Multiple} & \multicolumn{1}{c|}{\begin{tabular}[c]{@{}c@{}}Three\\ Layer\end{tabular}} & \multicolumn{1}{c|}{Multiple} & \multicolumn{1}{c|}{$\times$} & \multicolumn{1}{c|}{NA} & \multicolumn{1}{c|}{\begin{tabular}[c]{@{}c@{}}System \\Run Cost,\\ System \\ Utilization\end{tabular}} & \multicolumn{1}{c|}{Deadline} & \multicolumn{1}{c|}{NA} & \multicolumn{1}{c|}{Heuristic} & \multicolumn{1}{c|}{Low} & \multicolumn{1}{c|}{Low} & \multicolumn{1}{c|}{Low} & \multicolumn{1}{c|}{\begin{tabular}[c]{@{}c@{}}Scheduling realtime tasks in Fog networks\\ while considering task's security\end{tabular}} & \multicolumn{1}{c|}{\begin{tabular}[c]{@{}c@{}}Tag assignment for each service to show security\\intensiveness and to each resource to show their\\ trust level\end{tabular}} & \begin{tabular}[c]{@{}c@{}}Success\\ Ratio,\\ Throughput\end{tabular} \\ \hline

\multicolumn{1}{|c|}{\cite{zhang2024security}} & \multicolumn{1}{c|}{Task} & \multicolumn{1}{c|}{Multiple} & \multicolumn{1}{c|}{Independent} & \multicolumn{1}{c|}{\checkmark} & \multicolumn{1}{c|}{Multiple} & \multicolumn{1}{c|}{\begin{tabular}[c]{@{}c@{}}Three\\ Layer\end{tabular}} & \multicolumn{1}{c|}{Multiple} & \multicolumn{1}{c|}{\checkmark} & \multicolumn{1}{c|}{MDP} & \multicolumn{1}{c|}{\begin{tabular}[c]{@{}c@{}}Time,\\ Energy\end{tabular}} & \multicolumn{1}{c|}{Deadline} & \multicolumn{1}{c|}{Edge} & \multicolumn{1}{c|}{\begin{tabular}[c]{@{}c@{}}DRL\\ (DQN)\end{tabular}} & \multicolumn{1}{c|}{Low} & \multicolumn{1}{c|}{Mid} & \multicolumn{1}{c|}{Low} & \multicolumn{1}{c|}{\begin{tabular}[c]{@{}c@{}}Secure resource allocation in multi-Cloud\\ Edge computing minimizing\\ weighted delay and energy consumption\end{tabular}} & \multicolumn{1}{c|}{\begin{tabular}[c]{@{}c@{}}Hybrid AES-RSA encryption with MD5\\ integrity verification for data transmission\\ security in MEC-to-CC offloading\end{tabular}} & \begin{tabular}[c]{@{}c@{}}Response Time,\\ Energy\end{tabular} \\ \hline

\multicolumn{1}{|c|}{\cite{rahmani2024novel}} & \multicolumn{1}{c|}{Task} & \multicolumn{1}{c|}{Multiple} & \multicolumn{1}{c|}{Independent} & \multicolumn{1}{c|}{\checkmark} & \multicolumn{1}{c|}{Multiple} & \multicolumn{1}{c|}{\begin{tabular}[c]{@{}c@{}}Three\\ Layer\end{tabular}} & \multicolumn{1}{c|}{Multiple} & \multicolumn{1}{c|}{\checkmark} & \multicolumn{1}{c|}{MDP} & \multicolumn{1}{c|}{\begin{tabular}[c]{@{}c@{}}Time,\\ Energy\end{tabular}} & \multicolumn{1}{c|}{Deadline} & \multicolumn{1}{c|}{Edge} & \multicolumn{1}{c|}{\begin{tabular}[c]{@{}c@{}}DDRL\\ (A3C)\end{tabular}} & \multicolumn{1}{c|}{Low} & \multicolumn{1}{c|}{Mid} & \multicolumn{1}{c|}{Mid} & \multicolumn{1}{c|}{\begin{tabular}[c]{@{}c@{}}Multi-user task deployment optimization\\ in blockchain-enabled MEC-IoT networks\\ balancing energy, latency and security\end{tabular}} & \multicolumn{1}{c|}{\begin{tabular}[c]{@{}c@{}}Blockchain-based secure transaction\\ recording and immutable audit trail\\ for data integrity and trust\end{tabular}} & \begin{tabular}[c]{@{}c@{}}Response Time,\\ Energy\end{tabular} \\ \hline

\multicolumn{1}{|c|}{\cite{ebrahim2023privacy}} & \multicolumn{1}{c|}{Task} & \multicolumn{1}{c|}{Multiple} & \multicolumn{1}{c|}{Independent} & \multicolumn{1}{c|}{\checkmark} & \multicolumn{1}{c|}{Multiple} & \multicolumn{1}{c|}{\begin{tabular}[c]{@{}c@{}}Three\\ Layer\end{tabular}} & \multicolumn{1}{c|}{Multiple} & \multicolumn{1}{c|}{\checkmark} & \multicolumn{1}{c|}{MDP} & \multicolumn{1}{c|}{Time} & \multicolumn{1}{c|}{Privacy} & \multicolumn{1}{c|}{Edge} & \multicolumn{1}{c|}{\begin{tabular}[c]{@{}c@{}}DRL\\ (DDQN)\end{tabular}} & \multicolumn{1}{c|}{Low} & \multicolumn{1}{c|}{Mid} & \multicolumn{1}{c|}{Low} & \multicolumn{1}{c|}{\begin{tabular}[c]{@{}c@{}}Privacy-aware load balancing in Fog\\ networks minimizing waiting delay\\ without revealing Fog node information\end{tabular}} & \multicolumn{1}{c|}{\begin{tabular}[c]{@{}c@{}}Information hiding approach avoiding\\ disclosure of Fog load and resource\\ information to maintain provider privacy\end{tabular}} & \begin{tabular}[c]{@{}c@{}}Response Time\end{tabular} \\ \hline

\multicolumn{1}{|c|}{\cite{sun2024secure}} & \multicolumn{1}{c|}{Task} & \multicolumn{1}{c|}{Multiple} & \multicolumn{1}{c|}{Independent} & \multicolumn{1}{c|}{\checkmark} & \multicolumn{1}{c|}{Multiple} & \multicolumn{1}{c|}{\begin{tabular}[c]{@{}c@{}}Three\\ Layer\end{tabular}} & \multicolumn{1}{c|}{Multiple} & \multicolumn{1}{c|}{×} & \multicolumn{1}{c|}{MDP} & \multicolumn{1}{c|}{\begin{tabular}[c]{@{}c@{}}Time,\\ Energy\end{tabular}} & \multicolumn{1}{c|}{Security} & \multicolumn{1}{c|}{Edge} & \multicolumn{1}{c|}{\begin{tabular}[c]{@{}c@{}}DRL\\ (DQN)\end{tabular}} & \multicolumn{1}{c|}{Low} & \multicolumn{1}{c|}{Mid} & \multicolumn{1}{c|}{Low} & \multicolumn{1}{c|}{\begin{tabular}[c]{@{}c@{}}Optimizing secure task deployment\\ decisions while protecting data\\ privacy during transmission\end{tabular}} & \multicolumn{1}{c|}{\begin{tabular}[c]{@{}c@{}}RSA asymmetric encryption with\\ MD5 hashing for data integrity\\ and privacy protection\end{tabular}} & \begin{tabular}[c]{@{}c@{}}Response Time,\\ Energy\end{tabular} \\ \hline

\multicolumn{1}{|c|}{\cite{mohammadi2025security}} & \multicolumn{1}{c|}{Task} & \multicolumn{1}{c|}{Multiple} & \multicolumn{1}{c|}{Independent} & \multicolumn{1}{c|}{\checkmark} & \multicolumn{1}{c|}{Multiple} & \multicolumn{1}{c|}{\begin{tabular}[c]{@{}c@{}}Three\\ Layer\end{tabular}} & \multicolumn{1}{c|}{Multiple} & \multicolumn{1}{c|}{$\times$} & \multicolumn{1}{c|}{MINLP} & \multicolumn{1}{c|}{\begin{tabular}[c]{@{}c@{}}Time,\\ Energy\end{tabular}} & \multicolumn{1}{c|}{Security} & \multicolumn{1}{c|}{Edge} & \multicolumn{1}{c|}{\begin{tabular}[c]{@{}c@{}}Metaheuristic\\ (NSGA-II)\end{tabular}} & \multicolumn{1}{c|}{Mid} & \multicolumn{1}{c|}{Low} & \multicolumn{1}{c|}{Mid} & \multicolumn{1}{c|}{\begin{tabular}[c]{@{}c@{}}Resource allocation considering security\\ breach costs\end{tabular}} & \multicolumn{1}{c|}{\begin{tabular}[c]{@{}c@{}}Multi-objective optimization with adaptive crypto\\algorithm selection based on security levels\end{tabular}} & \begin{tabular}[c]{@{}c@{}}Response Time,\\ Energy\end{tabular} \\ \hline

\multicolumn{1}{|c|}{\cite{du2025secure}} & \multicolumn{1}{c|}{Task} & \multicolumn{1}{c|}{Multiple} & \multicolumn{1}{c|}{Independent} & \multicolumn{1}{c|}{×} & \multicolumn{1}{c|}{Multiple} & \multicolumn{1}{c|}{\begin{tabular}[c]{@{}c@{}}Three\\ Layer\end{tabular}} & \multicolumn{1}{c|}{Multiple} & \multicolumn{1}{c|}{\checkmark} & \multicolumn{1}{c|}{MINLP} & \multicolumn{1}{c|}{\begin{tabular}[c]{@{}c@{}}Security,\\ Energy\end{tabular}} & \multicolumn{1}{c|}{Resource} & \multicolumn{1}{c|}{Edge} & \multicolumn{1}{c|}{\begin{tabular}[c]{@{}c@{}}Metaheuristic\\ (GA)\end{tabular}} & \multicolumn{1}{c|}{Mid} & \multicolumn{1}{c|}{Low} & \multicolumn{1}{c|}{Low} & \multicolumn{1}{c|}{\begin{tabular}[c]{@{}c@{}}Secure computation offloading in Ocean IoT \\while preventing data interception and \\protecting client privacy \end{tabular}} & \multicolumn{1}{c|}{\begin{tabular}[c]{@{}c@{}}Symmetric encryption algorithms with\\ secure key management and cryptographic\\ frameworks for data transmission security\end{tabular}} & Security \\ \hline

\multicolumn{1}{|c|}{\cite{thangaraj2024msco}} & \multicolumn{1}{c|}{Task} & \multicolumn{1}{c|}{Multiple} & \multicolumn{1}{c|}{Independent} & \multicolumn{1}{c|}{\checkmark} & \multicolumn{1}{c|}{Multiple} & \multicolumn{1}{c|}{\begin{tabular}[c]{@{}c@{}}Three\\ Layer\end{tabular}} & \multicolumn{1}{c|}{Multiple} & \multicolumn{1}{c|}{\checkmark} & \multicolumn{1}{c|}{MIP} & \multicolumn{1}{c|}{\begin{tabular}[c]{@{}c@{}}Time,\\ Energy\end{tabular}} & \multicolumn{1}{c|}{Security} & \multicolumn{1}{c|}{Edge} & \multicolumn{1}{c|}{\begin{tabular}[c]{@{}c@{}}Metaheuristic\\ (GA-PSO)\end{tabular}} & \multicolumn{1}{c|}{Mid} & \multicolumn{1}{c|}{Low} & \multicolumn{1}{c|}{Low} & \multicolumn{1}{c|}{\begin{tabular}[c]{@{}c@{}}Secure computation offloading in \\blockchain-enabled environments to \\prevent malicious node attacks\end{tabular}} & \multicolumn{1}{c|}{\begin{tabular}[c]{@{}c@{}}Blockchain technology with PoW consensus \\mechanism to maintain decentralized data \\integrity and prevent single point of failure \end{tabular}} & \begin{tabular}[c]{@{}c@{}}Response Time,\\ Energy\end{tabular} \\ \hline

\multicolumn{1}{|c|}{Our Work} & \multicolumn{1}{c|}{service} & \multicolumn{1}{c|}{Multiple} & \multicolumn{1}{c|}{Dependent} & \multicolumn{1}{c|}{\checkmark} & \multicolumn{1}{c|}{Multiple} & \multicolumn{1}{c|}{\begin{tabular}[c]{@{}c@{}}Three\\ Layer\end{tabular}} & \multicolumn{1}{c|}{Multiple} & \multicolumn{1}{c|}{\checkmark} & \multicolumn{1}{c|}{MDP} & \multicolumn{1}{c|}{\begin{tabular}[c]{@{}c@{}}Time,\\ Security\end{tabular}} & \multicolumn{1}{c|}{\begin{tabular}[c]{@{}c@{}}Security,\\ Deadline\end{tabular}} & \multicolumn{1}{c|}{Edge} & \multicolumn{1}{c|}{DDRL} & \multicolumn{1}{c|}{Low} & \multicolumn{1}{c|}{High} & \multicolumn{1}{c|}{High} & \multicolumn{1}{c|}{\begin{tabular}[c]{@{}c@{}}Jointly optimizing response time\\ and security for service placement\\ in heterogeneous Fog computing\\ environments\end{tabular}} & \multicolumn{1}{c|}{\begin{tabular}[c]{@{}c@{}}Security optimization combining\\ hard constraint satisfaction with\\ security score maximization via\\ penalty-based reward mechanism\end{tabular}} & \begin{tabular}[c]{@{}c@{}}Response Time,\\ Security\end{tabular}  \\ \hline
\multicolumn{20}{|c|}{\begin{tabular}[C]{@{}c@{}}QCMIP: Quadratically Constrained Mixed Integer Programming, ILP: Integer Linear Programming, MINLP: Mixed Integer Non-Linear Programming, MIQP: Mixed Integer Quadratic Programming,  MDP: Markov Decision Process, MIP: Mixed-Integer Programming, NA: Not Available, Heter: Heterogeneity, PSO: Particle Swarm Optimization, \\DDRL: Distributed Deep Reinforcement Learning, B\&B: Branch and Bound, CMP: Complexity, ADAP: Adaptability, SCAL: Scalability \end{tabular}} \\ \hline
\end{tabular}%
}
\end{table*}
\par
Table~\ref{tab:relatedwork} provides a comprehensive comparison of our work with existing literature across several key areas, including service properties (e.g., structure, number, dependencies, heterogeneity), environmental properties (e.g., devices number, hierarchy, resource distribution, and heterogeneity), problem formulation (mathematical modeling, parameters, constraints), decision engine properties (placement layer, solver, complexity, adaptability, scalability), security properties (security goals, mitigation approaches, mechanisms), and the evaluation metrics used.
\par
The most complex Fog environments involve a diverse mix of IoT devices, services with varying service requirements (often with interdependent services), heterogeneous FSs, multi-layered Fog computing environment with heterogeneous resources, and high dynamism. In these complex and stochastic settings, the service placement for diverse IoT applications to optimize the response time and security score becomes a critical yet challenging problem. While some existing works (e.g., \cite{mann2020secure,elgendy2019resource,javanmardi2023s,mann2021security,singh2019scheduling,zhang2024security,rahmani2024novel,ebrahim2023privacy,sun2024secure,mohammadi2025security,thangaraj2024msco}) define security scores for FSs during placement, the practicalities of defining such scores remain unclear. Additionally, traditional solver approaches for service placement like heuristics and metaheuristics (e.g., \cite{sun2022security,mann2020secure,elgendy2019resource,casola2020security,javanmardi2023s,mann2021security,wang2022design,singh2019scheduling,mohammadi2025security,du2025secure,thangaraj2024msco}) struggle with finding efficient solutions in these environments due to their high time complexity, limited adaptability, and limited scalability \cite{garaali2022learning}. Moreover, DRL techniques used for service placement problem often face practical deployment challenges due to high exploration costs and slow convergence, especially as the number of features, environmental complexity, and placement problem constraints increase \cite{wang2020fast,garaali2022learning}. To tackle these limitations, SPA-DDRL addresses security quantification through a practical three-tier hierarchical model that captures security at control, capability, and configuration levels while incorporating comprehensive response time modeling, overcoming the ambiguity in existing security scoring approaches, and enabling precise dual-objective optimization. The distributed broker-learner architecture enhances scalability and adaptability compared to traditional centralized methods, while LSTM integration and PER significantly reduce convergence time and exploration costs compared to conventional DRL approaches. Also, sophisticated off-policy correction mechanisms, including importance sampling and gradient clipping ensure stable learning despite policy divergence in distributed settings, making the framework particularly suitable for large-scale dynamic Fog computing deployments.
\begin{figure}[t]
	\centering 
	\includegraphics[width=\linewidth, height=8cm]{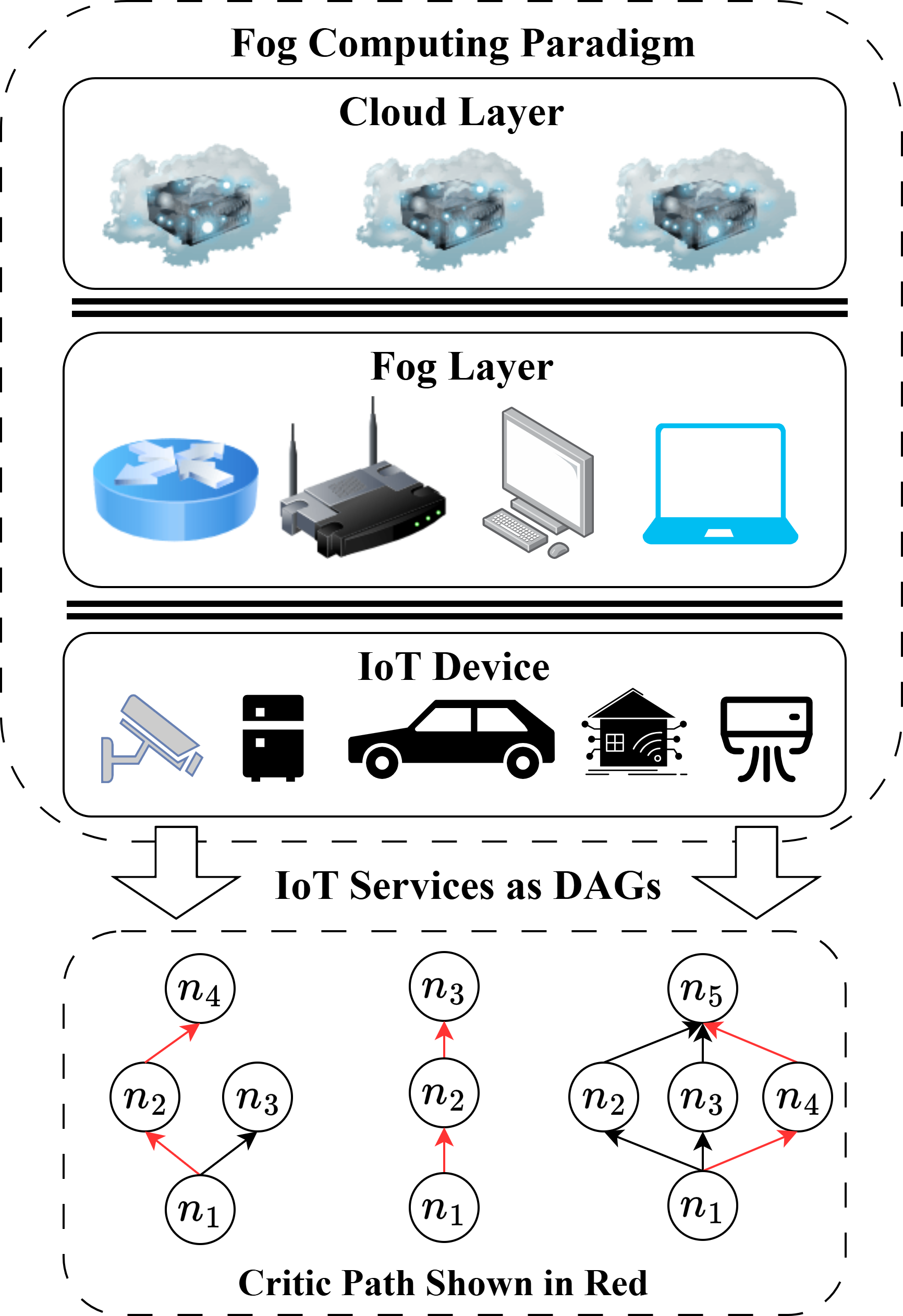}
	\caption{Three-tier Fog computing architecture, and a sample  service DAGs with critical paths highlighted in red}
	\label{fig:systemmodel}
\end{figure} 
%
%
\section{System Model and Problem Formulation}
\label{system}
We consider a fog computing system with three resource tiers, as shown in Figure~\ref{fig:systemmodel}: CSs, FSs, and IoT devices. CSs provide substantial computing power but suffer from high latency. FSs, distributed at the network edge, offer moderate resources with lower latency. IoT devices generate service requests but have limited local processing capacity. All parameters are summarized in Table 1 in the Appendix.
\subsection{Service Modeling}
A service is modeled as a DAG $\Gamma = (\mathcal{N}, \mathcal{A})$, where $\mathcal{N}$ denotes the set of tasks and $\mathcal{A}$ represents the set of directed edges between tasks. The service comprises $K$ tasks, i.e., $|\mathcal{N}| = K$, with each task indexed as $n_h$ where $h \in \{1, 2, ..., K\}$.

Each task $n_h$ is characterized by a five-tuple $\langle c_h, m_h, st_h, s_h, d_h \rangle$ where $c_h$ represents the required CPU cycles, $m_h$ specifies the memory demand, $st_h$ denotes the storage requirement, $s_h$ represents the security requirement, and $d_h$ indicates the deadline constraint.

A directed edge $a_{k,h} \in \mathcal{A}$ from task $n_k$ to task $n_h$ signifies that $n_h$ depends on the output of $n_k$. Each edge is associated with a weight $w_{k,h}$ where it quantifies the volume of data transmitted from $n_k$ to $n_h$. For task $n_h$, we define $\Pi(n_h) = \{n_k \mid a_{k,h} \in \mathcal{A}\}$ as the set of its immediate predecessor tasks.

\subsection{Problem Formulation}
Let $\mathcal{R}$ denote the set of available servers with cardinality $|\mathcal{R}| = R$. Each server $r^{p,q} \in \mathcal{R}$ is identified by type $p$ (e.g., CS, FS, IoT) and index $q$ within that type. A deployment scheme $\Phi$ shows the assignment of each task to one server:
\begin{equation}
    \Phi = \{\phi_{n_h} \mid n_h \in \mathcal{N}, \phi_{n_h} \in \mathcal{R}\}
\end{equation}
where $\phi_{n_h} = r^{p,q}$ shows task $n_h$ is assigned to server $r^{p,q}$. Each server is characterized by its processing capacity $\rho_{r^{p,q}}$, memory capacity $M_{r^{p,q}}$, storage capacity $ST_{r^{p,q}}$, physical location coordinates $(x_{r^{p,q}}, y_{r^{p,q}})$ in a two-dimensional space, and a set of enabled security configuration items $\mathcal{CNF}_{r^{p,q}}$ that determines its security capabilities' compliance.

\subsubsection{Response Time Model}
The completion time of task $n_h$ assigned to server $\phi_{n_h}$ consists of two components: computation time and data waiting time. Formally, we express this as:
\begin{equation}
    T_{\phi_{n_h}} = T_{\phi_{n_h}}^{\text{comp}} + T_{\phi_{n_h}}^{\text{wait}}
\end{equation}

The computation time is determined by the task's CPU requirement $c_h$ and the server's processing capacity $\rho_{\phi_{n_h}}$:
\begin{equation}
    T_{\phi_{n_h}}^{\text{comp}} = \frac{c_h}{\rho_{\phi_{n_h}}}
\end{equation}

The data waiting time represents the duration before all prerequisite inputs become available at the assigned server. For tasks with different server assignments, data transmission involves both bandwidth-constrained transfer and distance-dependent propagation. This is formulated as:
\begin{equation}
\scalebox{0.9}{$\displaystyle
T_{\phi_{n_h}}^{\text{wait}} = \max_{n_k \in \Pi(n_h)} \left[\left(\frac{w_{k,h}}{\omega(\phi_{n_k}, \phi_{n_h})} + \lambda(\phi_{n_k}, \phi_{n_h})\right) \cdot \delta_{\phi_{n_k}, \phi_{n_h}}\right]
$}
\end{equation}
where $\omega(\phi_{n_k}, \phi_{n_h})$ denotes the available bandwidth between the two servers, and $\lambda(\phi_{n_k}, \phi_{n_h})$ represents the propagation delay. The binary indicator $\delta_{\phi_{n_k}, \phi_{n_h}}$ takes value 1 if $\phi_{n_k} \neq \phi_{n_h}$ and 0 otherwise, thus eliminating network overhead for co-located tasks. The propagation delay is computed based on the Euclidean distance between server coordinates and the signal transmission speed $\nu$:
\begin{equation}
\lambda(\phi_{n_k}, \phi_{n_h}) = \frac{1}{\nu}\sqrt{(x_{\phi_{n_k}} - x_{\phi_{n_h}})^2 + (y_{\phi_{n_k}} - y_{\phi_{n_h}})^2}
\end{equation}

Due to the parallel execution capability inherent in DAG structures, the overall service latency is governed by the critical path $\mathcal{B}$, which represents the sequence of dependent tasks yielding the maximum cumulative execution time. We employ an upward ranking algorithm to identify critical path membership, which evaluates each task based on its computational cost and downstream dependencies. The critical path indicator $\beta_{n_h}$ is defined as:
\begin{equation}
    \beta_{n_h} = \begin{cases} 
        1, & n_h \in \mathcal{B} \\
        0, & \text{otherwise}
    \end{cases}
\end{equation}

Consequently, the total service response time under deployment $\Phi$ is computed as:
\begin{equation}
    \mathcal{L}(\Phi) = \sum_{h=1}^{K} \beta_{n_h} \cdot T_{\phi_{n_h}}
\end{equation}
\subsubsection{Security Model}
The open and distributed nature of Fog computing environments exposes them to diverse security threats, including data breaches, malicious attacks, and unauthorized access. Different IoT applications have varying security requirements, ranging from simple data integrity verification to complex end-to-end encryption and access control mechanisms. Simultaneously, heterogeneous computing resource providers possess different security capabilities and protection mechanisms. Therefore, we require a structured security model to precisely describe task security requirements, quantify resource provider security capabilities, and evaluate the matching degree between them.
Inspired by our previously patented solution in collaboration with Cisco Systems Australia \cite{patent}, we propose a three-tier hierarchical security model that provides fine-grained security requirement description and capability assessment framework from abstract security controls to specific configuration implementations. This model not only captures the complexity of real-world security requirements but also provides quantifiable security metrics for optimization algorithms.\par

\textbf{Hierarchical Security Architecture:} We assume a hierarchical Fog architecture and distributed brokers for placement decisions. The learner, brokers, and their communication channels are trustworthy, while FSs/CSs are untrusted and heterogeneous in security postures. Communication uses standard protocols, but channels may be insecure without enforced controls. Each task ($n_h$) is a primary asset with security requirements ($s_h$) that must be satisfied through appropriate server placement. Individual servers are modeled as having heterogeneous security capabilities and are inherently untrusted, requiring explicit checks to maintain compliance. The security framework validates their intrinsic Configuration Items (e.g., enabled crypto algorithms) against task requirements. 
\par
Our security model adopts a three-tier progressive structure: Controls (C), Capabilities (CP), and Configuration Items (CI).
The Security Control layer depicts high-level security objectives, such as Inventory of Authorized and Unauthorized Devices, and Data Encryption Protection \cite{sans2025ciscontrols}. Each security control corresponds to a specific security domain, reflecting particular security risks that the system needs to protect against. For task $n_h$ with security requirement $s_h$, it can be shown as a set of security controls:
\begin{equation}
s_h = \{C_k \mid k \in \mathcal{K}_h\}
\end{equation}
\noindent where $\mathcal{K}_h$ is the set of required security control indices for task $n_h$.
The Security Capability layer refines each security control into multiple specific security capabilities. For example, the Data Encryption Protection control may include capabilities such as Transmission Encryption, Storage Encryption, and Key Management. Each capability represents a specific technical dimension for implementing the security control. Different capabilities have varying importance in implementing security controls, so we assign weights to each capability, where the sum of all capability weights under the same control equals 1. Each control $C_k$ consists of multiple capabilities:
\begin{equation}
C_k = \{CP_l \mid l \in \mathcal{L}_{k}\}
\end{equation}
\noindent where $\mathcal{L}_{k}$ is the set of capability indices for control $C_k$.
The Configuration Item layer is the most specific implementation level, where each security capability consists of multiple configuration items. Configuration items represent whether specific security mechanisms are enabled, such as AES-256 Encryption Algorithm, Two-Factor Authentication, and Audit Logging. A configuration item value of 1 indicates that the security mechanism is deployed and functioning normally, while 0 indicates it is not deployed or unavailable. Each capability $CP_l$ comprises a set of configuration items:
\begin{equation}
CP_l = \{CI_i \mid i \in \mathcal{I}_l\}
\end{equation}
\noindent where $\mathcal{I}_l$ is the set of configuration item indices for capability $CP_l$, and each $CI_i \in \{0, 1\}$ is a binary indicator. 
\par
\textbf{Security Score Calculation:} We design a security scoring algorithm to quantify the matching degree between task requirements and resource capabilities.

For the capability $CP_l$ of task $n_h$ under deployment assignment $\phi_{n_h}$, we first calculate the configuration item satisfaction rate. Let $\mathcal{CNF}_{CP_l}$ denote the set of configuration items required by capability $CP_l$ (i.e., $\mathcal{CNF}_{CP_l} = \{CI_i \mid i \in \mathcal{I}_l\}$), and $\mathcal{CNF}_{\phi_{n_h}}$ be the set of configuration items supported by the assigned server. The configuration item satisfaction rate for capability $CP_l$ is computed by $SR_l$:
\begin{equation}
SR_l(\phi_{n_h}) = \frac{|\mathcal{CNF}_{CP_l} \cap \mathcal{CNF}_{\phi_{n_h}}|}{|\mathcal{CNF}_{CP_l}|} \times 100
\end{equation}

Next, we introduce a capability-level function to convert the configuration item satisfaction rate into a security score. The capability level function maps the satisfaction rate to discrete security levels, creating a non-linear transformation that emphasizes complete compliance:
\begin{equation}
\label{eq:capl}
\scalebox{0.75}{$\displaystyle
CapLevel(SR_l(\phi_{n_h})) = \begin{cases}
\sigma_{min}, & \text{if } SR_l(\phi_{n_h}) = \tau_{min} \\
f(\sigma, \tau, SR_l(\phi_{n_h})), & \text{if } \tau_{min} < SR_l(\phi_{n_h}) < \tau_{max} \\
\sigma_{max}, & \text{if } SR_l(\phi_{n_h}) = \tau_{max}
\end{cases}
$}
\end{equation}
\noindent where $\sigma_{min}$ and $\sigma_{max}$ represent minimum and maximum security scores, $\tau_{min}$ and $\tau_{max}$ are the boundary satisfaction rates, and $f(\sigma, \tau, SR_{l}(\phi_{n_h}))$ is a discretization function that maps intermediate satisfaction rates to appropriate security levels based on predefined thresholds. 

For control $C_k$ required by task $n_h$, we calculate the control-level security score by weighted aggregation of all capability scores it contains using $\mathcal{G}_k$ as shown below:
\begin{equation}
\mathcal{G}_{k}(\phi_{n_h}) = \frac{\sum_{l \in \mathcal{L}_k} w_l^{cp} \times CapLevel(SR_{l}(\phi_{n_h}))}{\sum_{l \in \mathcal{L}_k} w_l^{cp}}
\end{equation}
\noindent where $w_l^{cp}$ is the weight of capability $CP_l$ within control $C_k$.

Also, the control-level function, which maps control scores to discrete security levels is calculated as:
\begin{equation}
\label{eq:ctrl}
\scalebox{0.9}{$\displaystyle
CtrlLevel(\mathcal{G}_{k}(\phi_{n_h})) = \begin{cases}
\kappa_{min}, & \text{if } \mathcal{G}_{k}(\phi_{n_h}) = \xi_{min} \\
\kappa_{partial}, & \text{if } \xi_{min} < \mathcal{G}_{k}(\phi_{n_h}) < \xi_{max} \\
\kappa_{max}, & \text{if } \mathcal{G}_{k}(\phi_{n_h}) = \xi_{max}
\end{cases}
$}
\end{equation}
\noindent where $\kappa_{min}$, $\kappa_{partial}$, and $\kappa_{max}$ show non-compliant, partially compliant, and fully compliant security scores, respectively, while $\xi_{min}$ and $\xi_{max}$ define the boundary control score values. 

Finally, the overall security score for task $n_h$ under deployment assignment $\phi_{n_h}$ is obtained through weighted aggregation of all relevant security controls:
\begin{equation}
\mathcal{S}(\phi_{n_h}) = \frac{\sum_{k \in \mathcal{K}_h} w_k^{c} \times CtrlLevel(\mathcal{G}_{k}(\phi_{n_h}))}{\sum_{k \in \mathcal{K}_h} w_k^{c}}
\end{equation}
\noindent where $w_k^{c}$ is the weight of control $C_k$, reflecting its importance in the overall security assessment.

\textbf{Hard and Soft Constraint Handling:} In practical applications, security controls can be categorized into two types based on their criticality: hard constraints and soft constraints. Hard constraints represent critical security controls that must be strictly satisfied to ensure the fundamental security requirements of the application. These controls are typically related to regulatory compliance, data privacy protection, or mission-critical security policies. Soft constraints, on the other hand, represent desirable security enhancements that can be traded off with other objectives, such as performance or cost, during the optimization process.

We define a set of hard constraint controls $\mathcal{H}$, which is a subset of all security controls, that should achieve full compliance whenever possible. To effectively handle hard constraint violations, we introduce a penalty mechanism at the task level. The security score for task $n_h$ under deployment assignment $\phi_{n_h}$ is calculated as:
\begin{equation}
\scalebox{0.75}{$\displaystyle
\mathcal{S}(\phi_{n_h}) = \begin{cases}
\frac{\sum_{k \in \mathcal{K}_h} w_k^{c} \times CtrlLevel(\mathcal{G}_{k}(\phi_{n_h}))}{\sum_{k \in \mathcal{K}_h} w_k^{c}}, & \text{if } \forall k \in (\mathcal{K}_h \cap \mathcal{H}), \\
& \hspace{1em} CtrlLevel(\mathcal{G}_{k}(\phi_{n_h})) = 100 \\
P_{constraint}, & \text{otherwise}
\end{cases}
$}
\end{equation}
\noindent where $P_{constraint}$ is a large negative penalty value used to severely penalize hard constraint violations.
\par
The primary security objective of SPA-DDRL is guaranteed security compliance (Feasibility), defined by two sub-goals:
A) Quantitative compliance by maximizing the overall Service Security Score ($\mathcal{S}(\Phi)$) when ensuring a close match between the task's required security controls and the deployed resource's security capabilities, and B) Feasibility enforcement by strictly enforcing hard constraints ($\mathcal{H}$). Any service placement that violates a critical security control must be disqualified via a severe penalty in the objective function, thereby guaranteeing that the optimized solution is always securely viable.

Considering the security score of each task assignment within the service, the total service security score is:
\begin{equation}
\mathcal{S}(\Phi) = \frac{\sum_{h=1}^{K} \mathcal{S}(\phi_{n_h})}{K}
\end{equation}










%
\subsubsection{Optimization Problem}
The primary optimization goal is to minimize the response time of the service while maximizing the security score of the service placement. This is achieved by identifying the optimal deployment scheme for executing the service's tasks. To balance these potentially conflicting objectives, we employ a weighted multi-objective optimization approach with normalized objective components:
\begin{equation}
\scalebox{0.9}{$\displaystyle
\min \mathcal{W}(\Phi) = \min\Bigl(\alpha\,NORM_{\mathcal{L}}\!\bigl(\mathcal{L}(\Phi)\bigr) + \beta\,NORM_{\mathcal{S}}\!\bigl(\mathcal{S}(\Phi)\bigr)\Bigr)
$}
\end{equation}
where
\begin{equation}
NORM_{\mathcal{L}}(\mathcal{L}(\Phi)) = \frac{\mathcal{L}(\Phi) - \mathcal{L}_{min}}{\mathcal{L}_{max} - \mathcal{L}_{min}}
\end{equation}
\begin{equation}
NORM_{\mathcal{S}}(\mathcal{S}(\Phi)) = \frac{\mathcal{S}_{max} - \mathcal{S}(\Phi)}{\mathcal{S}_{max} - \mathcal{S}_{min}}
\end{equation}
In the objective function, $NORM_{\mathcal{L}}$ and $NORM_{\mathcal{S}}$ are the normalization functions for response time and security score, respectively, where $NORM_{\mathcal{S}}$ transforms the security score maximization problem into an equivalent minimization formulation. $\mathcal{L}_{min}$ and $\mathcal{L}_{max}$ represent the minimum and maximum possible response times, and $\mathcal{S}_{max}$ and $\mathcal{S}_{min}$ represent the maximum and minimum possible security scores. Through this normalization, both objective components are scaled to the same range, enabling the weight parameters $\alpha$ and $\beta$ to effectively balance the relative importance between response time and security objectives.
Our objective function minimizes the weighted objective:
\begin{equation}
\label{optimizationModel}
\min (\mathcal{W}(\Phi))
\end{equation}

Feasible deployments must satisfy several constraint categories. For allocation uniqueness ($CST1$), each task maps to exactly one server, enforced through:
\begin{eqnarray}
&&CST1:\;\sum_{r^{p,q} \in \mathcal{R}} \mathbb{I}_{\phi_{n_h} = r^{p,q}}=1,\; \forall n_h \in \mathcal{N}
\end{eqnarray}
where indicator $\mathbb{I}_{\phi_{n_h} = r^{p,q}}$ equals $1$ when task $n_h$ is assigned to server $r^{p,q}$ (i.e., $\phi_{n_h} = r^{p,q}$), and $0$ otherwise.

Resource capacity constraints prevent server oversubscription. Memory (CST2) and storage (CST3) capacity constraints ensure the total resource demands of all tasks assigned to each server remain within its physical capacity:
\begin{eqnarray}
&&CST2:\; \sum_{n_h \in \mathcal{N}} \mathbb{I}_{\phi_{n_h} = r^{p,q}} \times m_h  \leq M_{r^{p,q}},\; \forall r^{p,q} \in \mathcal{R}\\
&&CST3:\; \sum_{n_h \in \mathcal{N}} \mathbb{I}_{\phi_{n_h} = r^{p,q}} \times st_h  \leq ST_{r^{p,q}},\; \forall r^{p,q} \in \mathcal{R}
\end{eqnarray}

Deadline constraints ($CST4$) ensure that each task completes within its specified time bound:
\begin{eqnarray}
&&CST4:\; T_{\phi_{n_h}} \leq d_h,\; \forall n_h \in \mathcal{N}
\end{eqnarray}

Weight normalization constraints enforce proper weighting distributions across the security hierarchy and the objective function. $CST5$ ensures capability weights within each security control sum to 1, $CST6$ ensures control weights for each task sum to 1, and $CST7$ ensures the sum of response time and security score weight equals 1:
\begin{eqnarray}
&&CST5:\; \sum_{l \in \mathcal{L}_k} w_l^{cp} = 1,\; \forall k \in \mathcal{K}_h, \forall n_h \in \mathcal{N}\\
&&CST6:\; \sum_{k \in \mathcal{K}_h} w_k^{c} = 1,\; \forall n_h \in \mathcal{N}\\
&&CST7:\; \alpha + \beta = 1,\; \alpha, \beta \geq 0
\end{eqnarray}

The optimization problem addresses deployment decisions subject to multiple operational constraints. Given the exponential growth in solution space as server and task count increase, this problem belongs to the NP-hard class \cite{goudarzi2021distributedDDRL}, where polynomial-time optimal solutions are infeasible.
\section{Deep Reinforcement Learning Model}
\label{sec:DRLModel}
The service placement problem requires making a sequence of interdependent decisions, where each task assignment affects both immediate performance and future placement options. This sequential decision-making under uncertainty is naturally formulated as a Markov Decision Process (MDP), defined by the tuple $\langle\mathbb{S},\mathbb{A},\mathbb{P},\mathbb{R},\gamma\rangle$. 

At each decision epoch $t$, the agent observes state $s_t \in \mathbb{S}$ characterizing the current task and system conditions, executes action $a_t \in \mathbb{A}$ to select a server, and transitions to state $s_{t+1}$ with probability $\mathbb{P}(s_{t+1}|s_t, a_t)$ while receiving reward $r_t = \mathbb{R}(s_t, a_t)$. The discount factor $\gamma \in [0,1]$ balances immediate versus future rewards. The agent learns a policy $\pi: \mathbb{S} \rightarrow \mathbb{A}$ that maps states to actions, aiming to maximize the expected return:
\begin{equation}
J(\pi) = \mathbb{E}_{\pi} \left[\sum_{t=0}^{|\mathcal{N}|-1}\gamma^{t}r_t \mid s_0\right]
\end{equation}

The key components of our DRL formulation for security-aware service placement are defined as follows:

\begin{itemize}
\item \textbf{State Space $\mathbb{S}$:} At each time step $t \in \mathbb{T}$, the state $s_t \in \mathbb{S}$ captures both the current task to be placed and the dynamic state of all servers:
\begin{equation}
s_t = (F^{\mathcal{N}}_{t}, F^{\mathcal{R}}_{t})
\label{System_State}
\end{equation}
\noindent where $F^{\mathcal{N}}_{t} = \{f_i^{n_h} \mid 1 \leq i \leq k_1\}$ represents the feature vector of task $n_h$ being placed (including CPU cycles $c_h$, memory $m_h$, storage $st_h$, security requirements $s_h$, and data dependencies), and $F^{\mathcal{R}}_{t} = \{f_j^{r^{p,q}} \mid r^{p,q} \in \mathcal{R}, 1 \leq j \leq k_2\}$ encodes the feature vectors of all $R$ servers (including processing capacity, memory/storage utilization, bandwidth, location, and security capabilities).

\item \textbf{Action Space $\mathbb{A}$:} The action space corresponds to the set of available servers $\mathbb{A} = \mathcal{R}$. At each decision step $t$, the agent chooses action $a_t \in \mathbb{A}$ to assign the current task $n_h$ to a specific server, formally expressed as:
\begin{equation}
a_t: n_h \mapsto r^{p,q}, \quad \text{where } r^{p,q} \in \mathcal{R}
\end{equation}

\item \textbf{Reward Function $\mathbb{R}$:} Since the DRL agent makes sequential placement decisions for individual tasks, we define a task-level reward function that decomposes the service-level objective $\mathcal{W}(\Phi)$ from Equation~\ref{optimizationModel}. Through sequential decision-making, the agent learns to optimize task placements while considering their cumulative impact on service performance and security via the state representation that encodes previous placement decisions. The reward function incorporates deadline constraints through a penalty mechanism:
\begin{equation}
\label{eq.rewardFunctionTimeStep}
r_t = \begin{cases}
-\mathcal{W}(\phi_{n_h}), & \text{if } T_{\phi_{n_h}} \leq d_h \\
P_{failure}, & \text{otherwise}
\end{cases}
\end{equation}
\noindent where $T_{\phi_{n_h}}$ and $d_h$ are task completion time and deadline constraint. The negative sign transforms the minimization objective into a reward maximization problem. The large penalty $P_{failure}$ guides the agent away from deadline violations, ensuring feasible solutions.

\end{itemize}

The MDP formulation captures the essential properties of our security-aware service placement problem while providing a principled foundation for applying DRL techniques. The state representation encompasses both task requirements and provider capabilities, action space enables flexible resource selection, and reward function directly corresponds to the optimization objective, ensuring consistency between the learning process and the optimization outcome.

\section{SPA-DDRL: Distributed DRL Framework}
\label{sec:placement}

In this section, we describe the SPA-DDRL framework, the Security and Performance Aware Distributed Deep Reinforcement Learning framework for high-throughput service placement in Fog computing environments. SPA-DDRL integrates the Actor-Critic architecture with PER and LSTM networks to achieve efficient security-performance dual-objective optimization in highly heterogeneous and dynamic computing environments.

\begin{figure}[t]
\centering 
\includegraphics[width=1\linewidth, height=6cm]{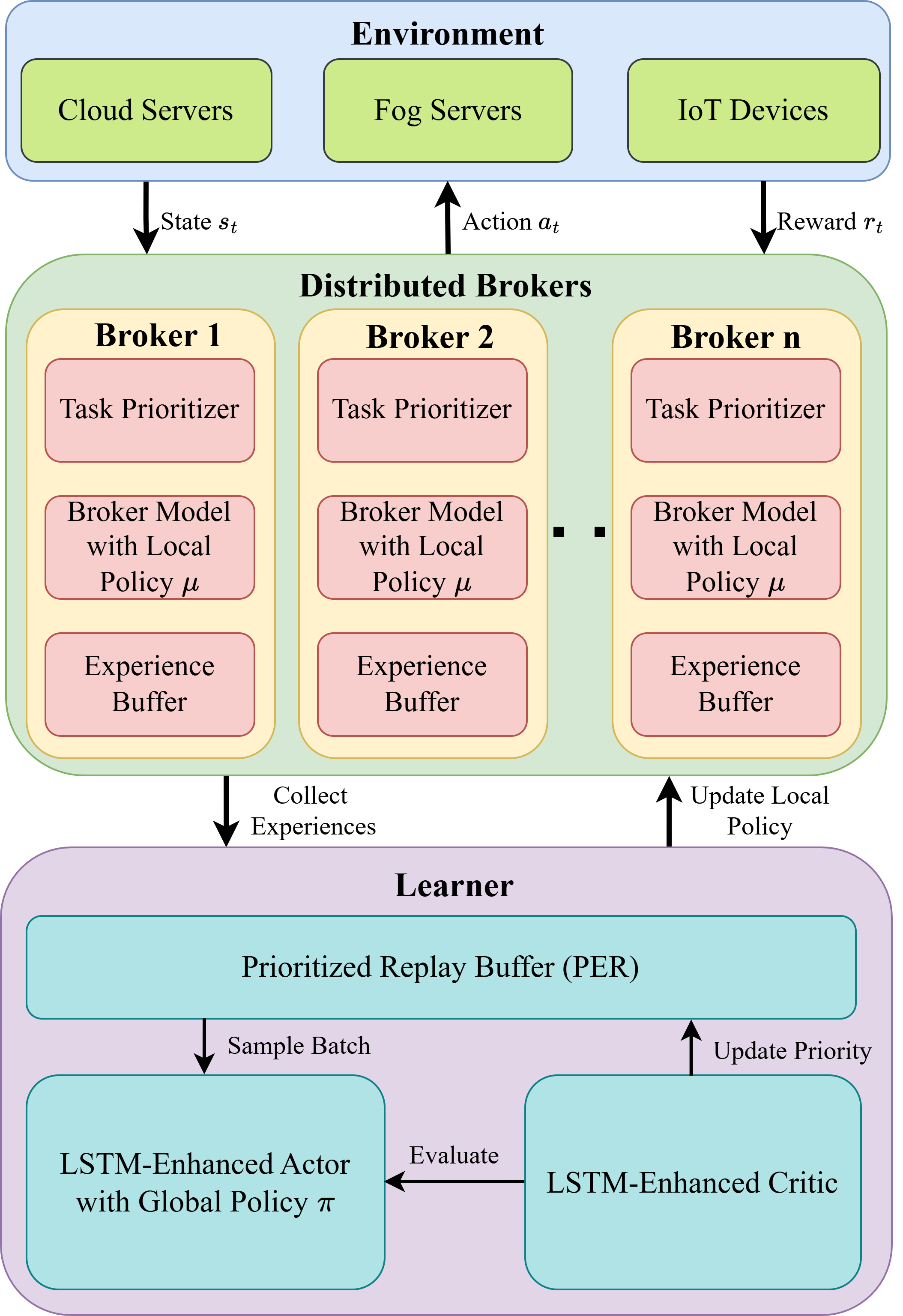}
\caption{An overview of SPA-DDRL framework}
\label{fig:spa-ddrl}
\end{figure}

SPA-DDRL employs a distributed broker-learner architecture where multiple intelligent brokers are deployed across heterogeneous servers to handle local placement decisions, while a centralized learner coordinates global policy optimization. Each broker utilizes LSTM-enhanced actor networks for real-time decision making, while the learner maintains both actor and critic networks with LSTM components for comprehensive policy learning and value estimation. This distributed design enables scalable security-performance optimization across heterogeneous Fog environments where security requirements and performance demands vary dynamically. Figure~\ref{fig:spa-ddrl} presents an overview of the SPA-DDRL framework. In what follows, we detail the operational mechanisms of brokers and the learner.

\subsection{Distributed Broker Operations}

Distributed brokers in SPA-DDRL function as autonomous decision-making entities positioned throughout the Fog infrastructure. Each broker processes incoming service requests and generates security-aware placement decisions using locally maintained LSTM-enhanced actor networks. The distributed deployment enables parallel processing of multiple service requests while maintaining security compliance and performance objectives. Algorithm~\ref{alg:broker-ops} outlines the broker operational procedure.

Each broker maintains a local policy $\mu$ that is periodically synchronized with the learner's global policy $\pi$ to ensure consistency across the distributed system (line 3). For each decision epoch, brokers process up to $N_{steps}$ placement decisions before synchronizing with the learner. When a new service arrives, the broker retrieves service metadata from the request queue $Q_R$, including security constraints, performance requirements, and task dependencies (line 7). The \textit{TaskPrioritizer()} function analyzes the service DAG structure and generates an ordered task sequence $\Psi_G$ using security-aware scheduling that prioritizes both critical path tasks and security-sensitive operations (line 8).

The \textit{StateComposer()} function constructs the decision state $s_t$ by combining server capability vectors $F^{\mathcal{R}}_{t}$, current task features $F^{\mathcal{N}}_{t}$, and security compliance indicators (line 9). This state representation is then processed by the \textit{DecisionEngine()}, which employs the LSTM-enhanced local policy $\mu$ to generate placement action $a_t$ (line 13). The \textit{RewardEvaluator()} computes the immediate reward $r_t$ based on the dual-objective function defined in Equation~\ref{eq.rewardFunctionTimeStep} (line 14). Experience tuples $(s_t, a_t, r_t, s_{t+1})$ are prioritized using the \textit{PriorityAssigner()} function, which considers both TD error magnitude and domain-specific factors such as security violations (line 16). Completed experiences are accumulated in the local experience buffer $EB$ until synchronization with the learner occurs (line 17).

\begin{algorithm}[!t]
\footnotesize
\caption{SPA-DDRL Broker Operations} \label{alg:broker-ops}
\SetKwData{Left}{left}
\SetKwData{This}{this}
\SetKwData{Up}{up}
\SetKwFunction{Union}{Union}
\SetKwFunction{FindCompress}{FindCompress}
\SetKwInOut{Input}{Input}
\SetKwInOut{Output}{Output}
\SetKwInOut{Parameter}{Parameter}
\Input{$\pi$: Global policy from learner}
\tcc{$N_{steps}$: Decision steps per epoch, $\mu$: Local broker policy, $EB$: Experience buffer, $Q_R$: Request queue, $G$: Service instance}
$status_{new} \leftarrow True$\\
\While {$True$}{
    $\mu \leftarrow$ SynchronizePolicy($\pi$)\\
    $step \leftarrow 0$ \\
    \While{$step < N_{steps}$}{
        \If{$status_{new} = True$}{
            $G \leftarrow Q_R$.dequeue()\\
            $\Psi_G \leftarrow$ TaskPrioritizer($G$)\\
            $s_t \leftarrow$ StateComposer($G$, $\mathcal{R}$, $\Psi_G$)\\
            $status_{new} \leftarrow False$\\
        }
        $s_t \leftarrow$ StateNormalizer($s_t$)\\
        $a_t \leftarrow$ DecisionEngine($s_t$, $\mu$) \% LSTM-enhanced placement decision\\
        $r_t \leftarrow$ RewardEvaluator($s_t$, $a_t$) \% $\rightarrow$ Eq.~\ref{eq.rewardFunctionTimeStep}\\
        $s_{t+1} \leftarrow$ StateTransition($s_t$, $a_t$)\\
        $p_t \leftarrow$ PriorityAssigner($s_t$, $a_t$, $r_t$, $s_{t+1}$) \% Priority calculation\\
        $EB$.store($s_t$, $a_t$, $r_t$, $s_{t+1}$, $p_t$)\\
        \If{ServiceComplete($G$)}{
            ServiceEvaluator($G$) \\
            $status_{new} \leftarrow True$\\
        }
        $step \leftarrow step + 1$\\
    }
    TransmitExperiences($EB$) \% Send prioritized experiences to learner\\
}
\end{algorithm}

\subsection{Centralized Learning Coordination}

The learner in SPA-DDRL orchestrates global policy optimization by aggregating prioritized experiences from distributed brokers and updating both actor and critic networks. The learner employs advanced off-policy correction techniques to handle the inherent policy lag between broker executions and learner updates. Algorithm~\ref{alg:learner-coord} details the learning coordination process.

\textbf{Off-Policy Correction Mechanisms:} The distributed nature of SPA-DDRL introduces policy divergence between broker policies $\mu$ and the learner's target policy $\pi$ \cite{espeholt2018impala}. To address this challenge, we employ a combination of importance sampling correction and gradient clipping. The importance sampling mechanism adjusts for the policy gap \cite{espeholt2018impala}, while gradient clipping prevents destructive parameter updates that could destabilize learning \cite{schulman2017proximal}.

The learner, as shown in Algorithm~\ref{alg:learner-coord}, continuously receives prioritized experience batches from active brokers and maintains a centralized replay buffer $RB$ (line 4). When sufficient experiences accumulate ($|RB| \geq BatchSize$), the \textit{SampleGenerator()} function creates training batches $SB$ using prioritized sampling (lines 5-7). The \textit{NetworkOptimizer()} function (begins at line 10) performs gradient-based updates. For each experience tuple $(s_j, a_j, r_j, s_{j+1}) \in SB$, the algorithm computes the importance sampling ratio (line 12):

\begin{equation}
\label{eq:importance-ratio}
\chi_j = \frac{\pi_{\theta}(a_j|s_j)}{\mu(a_j|s_j)}
\end{equation}

\noindent This ratio corrects for the policy divergence between broker execution policy $\mu$ and learner target policy $\pi_{\theta}$. Next, the corrected TD error is calculated in line 13:

\begin{equation}
\label{eq:td-correction}
\psi_j = \varrho_j \left(r_j + \gamma V_{\phi}(s_{j+1}) - V_{\phi}(s_j)\right)
\end{equation}

\noindent where $\varrho_j = \min(\bar{\varrho}, \chi_j)$ is the clipped importance weight and $V_{\phi}(s_j)$ represents the critic network's state value estimation.

The advantage estimation incorporates temporal dependencies and is computed in line 14:

\begin{equation}
\label{eq:advantage-est}
\hat{V}_j = \sum_{k=0}^{H-1} (\tau \gamma)^k \left(\prod_{l=0}^{k-1} \sigma_l\right) \psi_{j+k}
\end{equation}

\noindent where $\tau$ controls the bias-variance trade-off, $\gamma$ is the discount factor, and $\sigma_l$ are importance sampling clipping weights. After computing these values for all experiences in the batch, the policy gradient is calculated (line 16):

\begin{equation}
\label{eq:policy-gradient}
\nabla_{\theta} L_{\pi}(\theta) = \frac{1}{|SB|} \sum_{j \in SB} \min \left(\chi_j \cdot \hat{V}_j, \text{clip}(\chi_j, 1-\eta, 1+\eta) \cdot \hat{V}_j\right)
\end{equation}

\noindent where $\text{clip}(\cdot, 1-\eta, 1+\eta)$ is the clipping operation with threshold $\eta$ that ensures stable policy updates. The critic network gradient is computed in line 17:

\begin{equation}
\label{eq:critic-loss}
\nabla_{\phi} L_V(\phi) = \frac{1}{|SB|} \sum_j \left(V_{\phi}(s_j) - \hat{R}_j\right) \nabla_{\phi} V_{\phi}(s_j)
\end{equation}

\noindent where $\phi$ and $\hat{R}_j$ show critic parameters and target value. The network parameters are then updated using the computed gradients: actor parameters $\theta$ (line 18) and critic parameters $\phi$ (line 19). Finally, the \textit{PriorityUpdater()} function adjusts experience priorities based on updated TD errors (line 20), and the learner redistributes the updated global policy to all active brokers through \textit{BroadcastPolicy()} (line 21).

\begin{algorithm}[!t]
\footnotesize
\caption{SPA-DDRL Learning Coordination} \label{alg:learner-coord}
\SetKwData{Left}{left}
\SetKwData{This}{this}
\SetKwData{Up}{up}
\SetKwFunction{Union}{Union}
\SetKwFunction{FindCompress}{FindCompress}
\SetKwInOut{Input}{Input}
\SetKwInOut{Output}{Output}
\SetKwInOut{Parameter}{Parameter}
\Input{$EB_i$: Experience batches from brokers}
\tcc{$B$: Broker list, $\pi$: Global policy, $RB$: Replay buffer, $SB$: Sample batch}
\While {$True$}{
    $ready \leftarrow False$, $RB \leftarrow \emptyset$\\
    \While{$ready = False$}{
        $RB$.append($EB_i$) \% Collect prioritized experiences\\
        \If{$|RB| \geq BatchSize$}{
            $SB \leftarrow$ SampleGenerator($RB$) \% Prioritized sampling\\
            $ready \leftarrow True$\\
        }
    }
    NetworkOptimizer($SB$):\\
    \Indp
    \ForEach{$(s_j, a_j, r_j, s_{j+1}) \in SB$}{
        $\chi_j \leftarrow \frac{\pi_{\theta}(a_j|s_j)}{\mu(a_j|s_j)}$ \% Importance sampling ratio\\
        $\psi_j \leftarrow \varrho_j (r_j + \gamma V_{\phi}(s_{j+1}) - V_{\phi}(s_j))$ \% TD error\\
        $\hat{V}_j \leftarrow \sum_{k=0}^{H-1} (\tau \gamma)^k (\prod_{l=0}^{k-1} \sigma_l) \psi_{j+k}$ \% Advantage\\
    }
    $\nabla_{\theta} L_{\pi}(\theta) \leftarrow \frac{1}{|SB|} \sum_{j \in SB} \min(\chi_j \cdot \hat{V}_j, \text{clip}(\chi_j, 1-\eta, 1+\eta) \cdot \hat{V}_j)$\\
    $\nabla_{\phi} L_V(\phi) \leftarrow \frac{1}{|SB|} \sum_j (V_{\phi}(s_j) - \hat{R}_j) \nabla_{\phi} V_{\phi}(s_j)$\\
    Update actor parameters: $\theta \leftarrow \theta + \alpha_{\pi} \nabla_{\theta} L_{\pi}(\theta)$\\
    Update critic parameters: $\phi \leftarrow \phi + \alpha_V \nabla_{\phi} L_V(\phi)$\\
    \Indm
    PriorityUpdater($SB$) \% Update experience priorities\\
    BroadcastPolicy($B$, $\pi$) \% Distribute updated policy\\
}

\end{algorithm}
\hspace{-0.5cm}
\subsection{Prioritized Experience Management}

SPA-DDRL implements a PER mechanism to improve learning efficiency by focusing on experiences that provide the greatest learning potential. In conventional experience replay, experiences are sampled uniformly, leading to inefficient learning from less informative transitions \cite{schaul2015prioritized}. PER addresses this by assigning priority weights to experiences based on their TD error magnitude.

Experience priority in SPA-DDRL is computed as:

\begin{equation}
\text{priority}(e_j) = |\psi_j| + \varepsilon
\end{equation}

\noindent where $\psi_j$ is the TD error from Equation~\ref{eq:td-correction} and $\varepsilon$ ensures non-zero sampling probability for all experiences. Sampling probabilities follow the prioritized distribution:

\begin{equation}
P_j = \frac{p_j^{\nu}}{\sum_k p_k^{\nu}}
\end{equation}

\noindent where $p_j$ is the priority of experience $j$ and $\nu$ controls prioritization intensity ($\nu = 0$ yields uniform sampling). To correct for sampling bias, importance sampling weights are applied during gradient computation:

\begin{equation}
w_j = \left(\frac{1}{N} \cdot \frac{1}{P_j}\right)^{\iota}
\end{equation}

\noindent where $N$ is the buffer capacity and $\iota$ controls bias correction strength, typically annealed from 0 to 1 during training.

\subsection{LSTM Enhancement in SPA-DDRL}

SPA-DDRL incorporates LSTM networks into the Actor-Critic architecture to capture temporal dependencies in the dynamic Fog environment. Unlike traditional feedforward networks, LSTM networks can maintain long-term memory, which is crucial for Fog environments where current service placement decisions affect future resource availability and security states. The LSTM enhancement enables the framework to model sequential decision-making patterns and learn from historical state-action correlations.

\textbf{LSTM-Enhanced Actor Network:} The actor policy integrates LSTM hidden states to maintain temporal context across placement decisions. The key advantage of this design lies in its ability to recognize periodic patterns such as temporal regularities in service requests, evolution trends of security threats, and variation patterns in resource utilization. At time step $t$, the LSTM-enhanced actor processes the current state representation and previous hidden state:

\begin{equation}
h_t^{\pi} = \text{LSTM}_{\theta}([F^{\mathcal{N}}_{t}, F^{\mathcal{R}}_{t}], h_{t-1}^{\pi})
\end{equation}

\noindent where $h_t^{\pi}$ is the actor's hidden state at time $t$, $\theta$ represents the actor network parameters, and $[F^{\mathcal{N}}_{t}, F^{\mathcal{R}}_{t}]$ is the concatenated state representation. By maintaining historical information, the actor can make more informed decisions, particularly in recognizing historical security compliance patterns for similar service types, temporal correlations between task placements and overall service performance, and long-term trends in resource availability and security capabilities. The policy distribution is then computed as:

\begin{equation}
\pi_{\theta}(a_t|s_t, h_t^{\pi}) = \text{softmax}(W_{\pi} h_t^{\pi} + b_{\pi})
\end{equation}

\noindent where $W_{\pi}$ and $b_{\pi}$ are learnable parameters mapping the hidden state to action probabilities.

\textbf{LSTM-Enhanced Critic Network:} The critic network employs LSTM components to estimate state values considering temporal context. This is particularly important for security-performance evaluation, as the value of a placement decision depends not only on immediate rewards but also on long-term security compliance and performance implications. The LSTM enables the critic to model the evolution of security threats that may affect long-term placement strategies, temporal dependencies between security control implementations and their effectiveness, dynamic changes in resource security capabilities over time, and correlation between historical placement decisions and long-term service performance. The critic's LSTM processes the same state inputs but maintains separate hidden states:

\begin{equation}
h_t^V = \text{LSTM}_{\phi}([F^{\mathcal{N}}_{t}, F^{\mathcal{R}}_{t}], h_{t-1}^V)
\end{equation}

\noindent where $h_t^V$ represents the critic's hidden state and $\phi$ denotes the critic parameters. The state value estimation incorporates temporal information:

\begin{equation}
V_{\phi}(s_t, h_t^V) = W_V h_t^V + b_V
\end{equation}

\textbf{Temporal Advantage Estimation:} The temporal context enables more accurate advantage estimation by considering historical reward patterns. This enhanced advantage computation better reflects the long-term impact of decisions, particularly in complex security-performance trade-off scenarios. The advantage function becomes:

\begin{equation}
A_t = \sum_{k=0}^{T-t} \gamma^k r_{t+k} + \gamma^{T-t+1} V_{\phi}(s_{T+1}, h_{T+1}^V) - V_{\phi}(s_t, h_t^V)
\end{equation}

\noindent where $\gamma$ is the discount factor and $T$ is the episode length.

\textbf{Gradient Computation with Temporal Dependencies:} The policy gradient integrates LSTM-based temporal dependencies via backpropagation through time, enabling the network to capture complex correlation patterns across successive time steps. This is crucial for modeling security threat evolution and resources' dynamic changes:

\begin{equation}
\nabla_{\theta} J(\theta) = \mathbb{E}_{\tau} \left[\sum_{t=0}^{T} \nabla_{\theta} \log \pi_{\theta}(a_t|s_t, h_t^{\pi}) A_t\right]
\end{equation}

\noindent where $\tau = \{s_0, a_0, r_0, \ldots, s_T, a_T, r_T\}$ represents the trajectory sequence.

The LSTM integration with PER creates enhanced learning efficiency where experiences with high temporal significance receive prioritized sampling, accelerating learning of complex temporal patterns in security-performance optimization. The distributed broker-learner architecture further amplifies these benefits by enabling parallel exploration of diverse security-performance conditions across the Fog infrastructure, leading to more robust policy learning.

\section{Performance Evaluation}
\label{sec:evaluation}
This section presents the experimental evaluation of SPA-DDRL for security-performance optimization in Fog computing environments. We establish the evaluation setup, introduce baseline techniques for comparison, configure hyperparameters, and analyze performance across multiple metrics to demonstrate the framework's effectiveness in balancing security compliance with performance objectives. 

\subsection{Evaluation Setup}
To validate the effectiveness of SPA-DDRL in heterogeneous Fog computing, we design and implement a simulation platform for multi-tier computing infrastructure. The platform constructs a distributed environment containing 100 servers, including 20 high-performance CSs, 30 medium-performance FSs, and 50 resource-constrained IoT devices.

\textbf{Computing Infrastructure Design:} Following \cite{10643348, li2024next}, CSs feature high-end hardware resources: CPU core count ranges from 4 to 32, with processing capability reaching 10,000-100,000 MIPS per core, memory capacity configured as 16-128 GB, and storage space of 500-10,000 GB. Following \cite{mahmud2022ifogsim2, nahar2024clouds}, FSs utilize moderate hardware configurations: CPU core count of 2-8, processing capability ranging from 5,000-20,000 MIPS, memory capacity of 4-32 GB, and storage space of 200-500 GB. Following \cite{mahmud2022ifogsim2, goudarzi2023mu}, IoT devices operate with resource-constrained terminals: CPU cores of 1-2, processing capability of 1,000-5,000 MIPS, memory of 1-2 GB, and storage of 10-100 GB.

\textbf{Network Connectivity Modeling:} The system adopts a fully connected network topology, precisely modeling communication capabilities between servers/devices through a bandwidth matrix. Following \cite{wang2025reinfog, zhang2025collaborative, chen2025revenue}, each server/device has different network interface capabilities based on its type: IoT devices support 10-50 Mbps bidirectional transmission, FSs support 50-200 Mbps, and CSs support 100-1,000 Mbps. 

\textbf{Multi-tier Security Configuration:} The security mechanism adopts a three-tier hierarchical structure: 15 categories of basic security controls \cite{sans2025ciscontrols}, each category containing 5 implementation capabilities, and each capability composed of 3 configuration items. Different types of servers/devices implement different subsets of the 15 security controls based on their computational capabilities and deployment requirements. For instance, resource-rich Cloud servers may implement comprehensive security controls, while resource-constrained IoT devices implement only essential security mechanisms, creating a heterogeneous security landscape across the Fog computing infrastructure. 

In our implementation, without loss of generalizability, the capability-level function (i.e., Equation \ref{eq:capl}) is defined as:
\begin{equation}
\scalebox{0.7}{$\displaystyle
CapLevel(SR_{l}(\phi_{n_h})) = \begin{cases}
0, & \text{if } SR_{l}(\phi_{n_h}) = 0 \\
25, & \text{if } 0 < SR_{l}(\phi_{n_h}) \leq 33 \\
50, & \text{if } 33 < SR_{l}(\phi_{n_h}) \leq 66 \\
75, & \text{if } 66 < SR_{l}(\phi_{n_h}) < 100 \\
100, & \text{if } SR_{l}(\phi_{n_h}) = 100
\end{cases}
$}
\end{equation}
Similarly, the control-level function (i.e., Equation \ref{eq:ctrl}) are:
\begin{equation}
CtrlLevel(\mathcal{G}_{k}(\phi_{n_h})) = \begin{cases}
0, & \text{if } \mathcal{G}_{k}(\phi_{n_h}) = 0 \\
50, & \text{if } 0 < \mathcal{G}_{k}(\phi_{n_h}) < 100 \\
100, & \text{if } \mathcal{G}_{k}(\phi_{n_h}) = 100
\end{cases}
\end{equation}

\textbf{Heterogeneous Service Workload Construction:} We design complex service models based on DAGs to simulate real computing workloads. Similar to \cite{wang2025tf, wang2020fast}, each service contains multiple interdependent tasks with different computational, communication, and security requirements. The service generation process is divided into two stages: topology structure construction and task attribute assignment. In the topology structure construction stage, we control the shape characteristics of DAGs through three key parameters: task count $K$ determines the scale complexity of services, set to 6 levels $K \in \{5, 10, 20, 40, 80, 100\}$; the $fat$ parameter controls the width-to-height ratio of DAGs, affecting the parallelization degree of tasks, set to 5 levels $fat \in \{0.2, 0.4, 0.6, 0.8, 1.0\}$, where smaller $fat$ values generate more sequential execution structures and larger $fat$ values generate more parallel execution structures; the $density$ parameter controls the connection density between tasks, affecting the complexity of dependency relationships, also set to 5 levels $density \in \{0.2, 0.4, 0.6, 0.8, 1.0\}$, where smaller $density$ values produce loosely coupled task structures and larger $density$ values produce tightly coupled task structures. Through the combination of these three parameters, we generate 5 different topology variants for each parameter configuration to ensure structural diversity. In the task attribute assignment stage, we assign specific resource requirements and constraints to each generated topology structure. Based on \cite{9573423, goudarzi2023mu}, for computational requirements, each task's CPU workload is set to 0.5-100 million instructions, with memory requirements ranging from 10-1,000 MB; for time constraints, task execution deadlines are set to 10-1,000 milliseconds; for communication requirements, data transmission between tasks is set to 1-1,000 KB; for security requirements, each task randomly selects required security services from 15 security control categories. To increase attribute configuration diversity, each topology structure is paired with 10 different attribute combinations, allowing the same dependency structure to have different resource and security characteristics. Through the above generation strategy, we construct a comprehensive dataset containing 7,500 heterogeneous service instances, covering various service patterns from simple linear tasks to complex parallel tasks, as well as different security levels from low security requirements to high security requirements, providing sufficient test scenarios for security-performance dual-objective optimization. 
\par
The following baseline techniques are implemented:
\begin{itemize}
\item X-DDRL: The extended version of the method presented in \cite{goudarzi2021distributedDDRL} is employed as a baseline, adopting IMPALA as the underlying DDRL framework. We updated the reward function to support security optimization.
\item A3C-AHP: The improved version of the method in \cite{rahmani2024novel} is employed. It is adapted for service placement in heterogeneous computing environments comprising multiple IoT devices, Fog servers, and Cloud servers. Also, the reward function is updated to support security and response time optimization for service placement. Its underlying DDRL framework is A3C.
\item DRLIS: The extended version of the method presented in \cite{wang2024deep} is employed as a baseline. The reward function is updated to support security optimization. This method adopts PPO as the DRL framework.
\item PARL: The improved version of the method presented in \cite{ebrahim2023privacy} is employed as a baseline. It is adapted for service placement in heterogeneous computing environments. The reward function is updated to support security and response time optimization. Its underlying DRL framework is based on DDQN.
\item SCRA: The enhanced version of the method in \cite{zhang2024security} is employed as a baseline. This technique is adapted for service placement, and the reward function is updated to optimize service security score and response time. This method uses DQN as the DRL framework.
\end{itemize}

\subsection{Hyperparameters Tuning}
SPA-DDRL employs identical two-layer fully connected neural networks across all distributed brokers with Tanh activation functions. Through grid search optimization, we configure the learning rate at 0.01 with Adam optimizer, discount factor $\gamma$ at 0.9, and batch size of 128. The V-trace importance sampling weights $\rho$ and $c$ are set to 1.0, clipping threshold $\epsilon$ to 0.2, and GAE lambda to 0.95. The experience buffer maintains 10,000 samples with gradient updates every 2 steps. Complete configurations are shown in Table~\ref{tab:hyperparameters}. For fair comparison, all baseline techniques are optimized through comprehensive grid search, with their configurations presented in Table~\ref{tab:baseline_hyperparameters}.

\begin{table}[!t]
\centering
\caption{Hyperparameters of SPA-DDRL}
\adjustbox{width=\columnwidth,center}{
\begin{tabular}{ll|ll}
\hline
\textbf{Parameter} & \textbf{Value} & \textbf{Parameter} & \textbf{Value} \\
\hline
FC layers & 2 & Learning Rate $lr$ & 0.01 \\
Gradient Steps & 2 & Discount Factor $\gamma$ & 0.9 \\
Optimization Technique & Adam & Batch Size & 128 \\
Activation Function & Tanh & Buffer Size & 10000 \\
Clipping Constant $\epsilon$ & 0.2 & V-trace $\rho$ & 1.0 \\
GAE Lambda $\lambda$ & 0.95 & V-trace $c$ & 1.0 \\
\hline
\end{tabular}
}
\label{tab:hyperparameters}
\end{table}

\begin{table}[!t]
\centering
\caption{Hyperparameters of baseline techniques}
\adjustbox{width=\columnwidth,center}{
\begin{tabular}{lccccc}
\hline
\textbf{Hyperparameters} & \textbf{X-DDRL} & \textbf{A3C-AHP} & \textbf{DRLIS} & \textbf{PARL} & \textbf{SCRA} \\
\hline
Fully Connected Layers & 2 & 2 & 2 & 2 & 2 \\
Activation Function & TanH & TanH & TanH & ReLU & ReLU \\
Learning Rate & 0.01 & 0.001 & 0.01 & 0.01 & 0.01 \\
Discount Factor $\gamma$ & 0.99 & 0.999 & 0.9 & 0.999 & 0.999 \\
Batch Size & 64 & 32 & 128 & 32 & 64 \\
Buffer Size & 100000 & N/A & N/A & 50000 & 50000 \\
\hline
\end{tabular}
}
\label{tab:baseline_hyperparameters}
\end{table}

\subsection{Performance Analysis}
To evaluate SPA-DDRL's effectiveness, we conduct a multifaceted performance analysis covering ablation studies, convergence behavior, scalability performance, training efficiency, and computational overhead. 
\par
\subsubsection{Ablation Analysis}
To evaluate the effectiveness of the key components in SPA-DDRL, we conduct an ablation analysis examining the individual contributions of LSTM and PER to the overall performance. For this experiment, we use services with varying complexity where the training dataset includes services with task counts $K \in \{5, 10, 20, 80, 100\}$, while the evaluation is performed on services with $K = 40$ tasks.

\begin{figure}[!t]
\centering
\includegraphics[width=1\columnwidth, height=4cm]{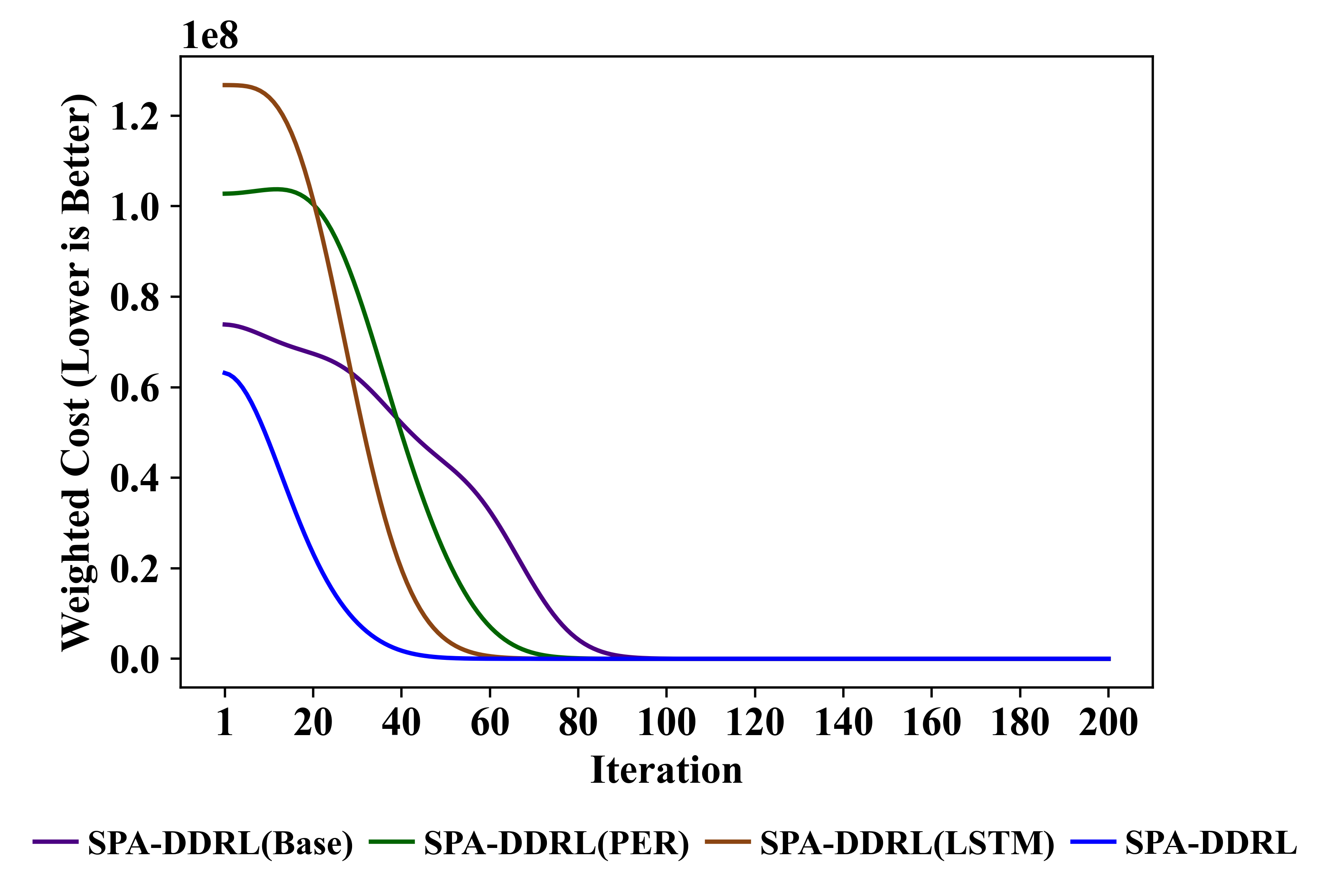}
\caption{Ablation analysis of SPA-DDRL variants}
\label{fig:ablation_analysis}
\end{figure}

Figure~\ref{fig:ablation_analysis} presents the convergence performance of different SPA-DDRL variants: SPA-DDRL(Base) without LSTM and PER, SPA-DDRL(PER) with only PER enabled, SPA-DDRL(LSTM) with only LSTM enabled, and the complete SPA-DDRL with both components. The results demonstrate that each component significantly contributes to faster convergence. SPA-DDRL achieves the fastest convergence around iteration 40, followed by SPA-DDRL(LSTM) at iteration around 60, SPA-DDRL(PER) at iteration around 70, and SPA-DDRL(Base) at iteration around 90. The LSTM component provides temporal dependency modeling that helps the agent better understand the dynamic Fog computing environment, resulting in approximately 33\% faster convergence compared to the base version. PER contributes by prioritizing important experiences during training, leading to more efficient learning and 22\% improvement in convergence speed. The combination of both components in the complete SPA-DDRL achieves synergistic effects, delivering the most efficient learning performance with 56\% improvement. In summary, both LSTM and PER are essential to the SPA-DDRL framework, jointly enabling superior convergence performance in complex, multi-objective service placement.

\begin{figure*}[!th]
\centering
\begin{subfigure}[b]{0.33\textwidth}
    \centering
    \includegraphics[width=\textwidth]{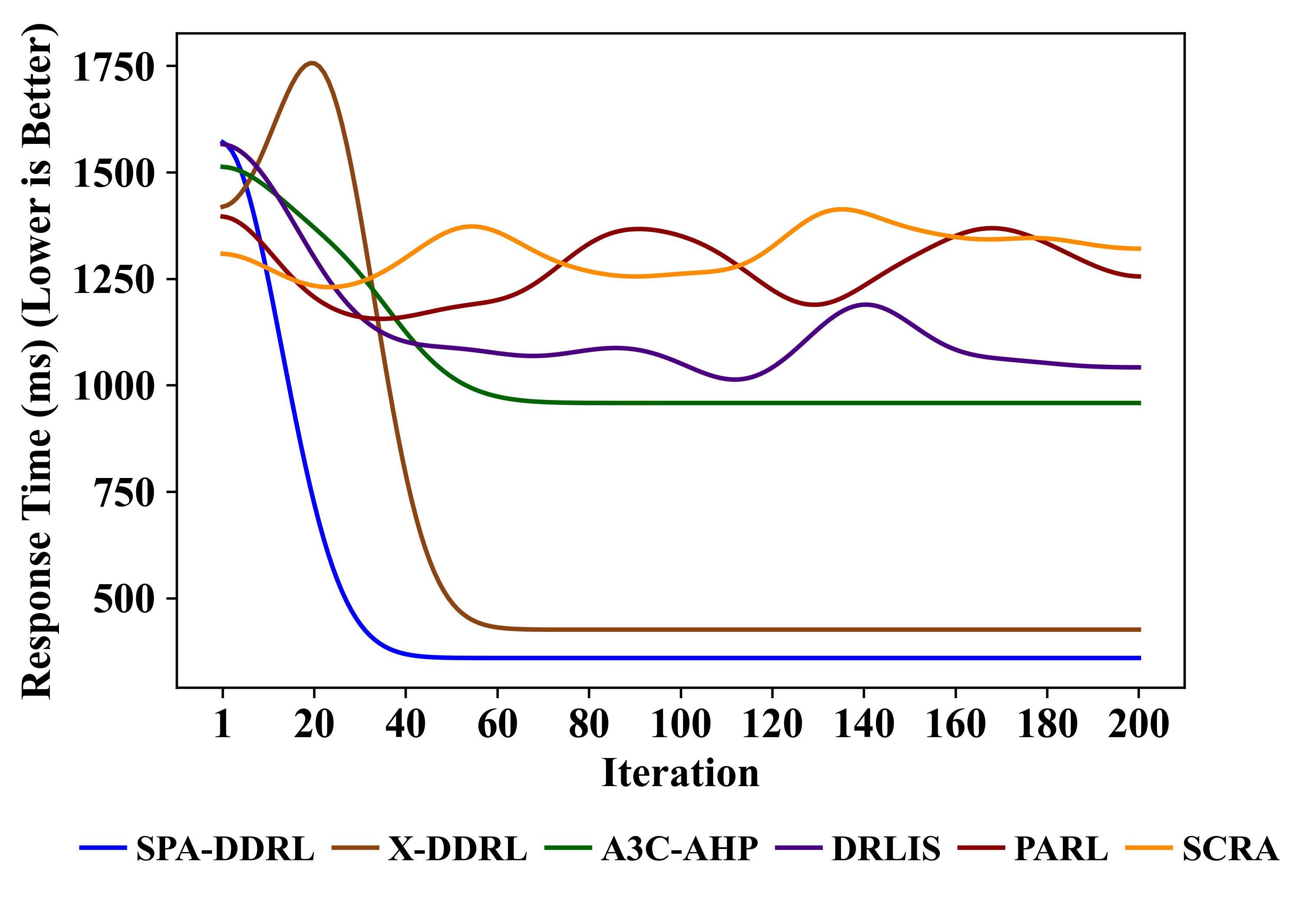}
    \caption{Response Time Convergence}
    \label{fig:response_time_convergence}
\end{subfigure}
\hfill
\begin{subfigure}[b]{0.33\textwidth}
    \centering
    \includegraphics[width=\textwidth]{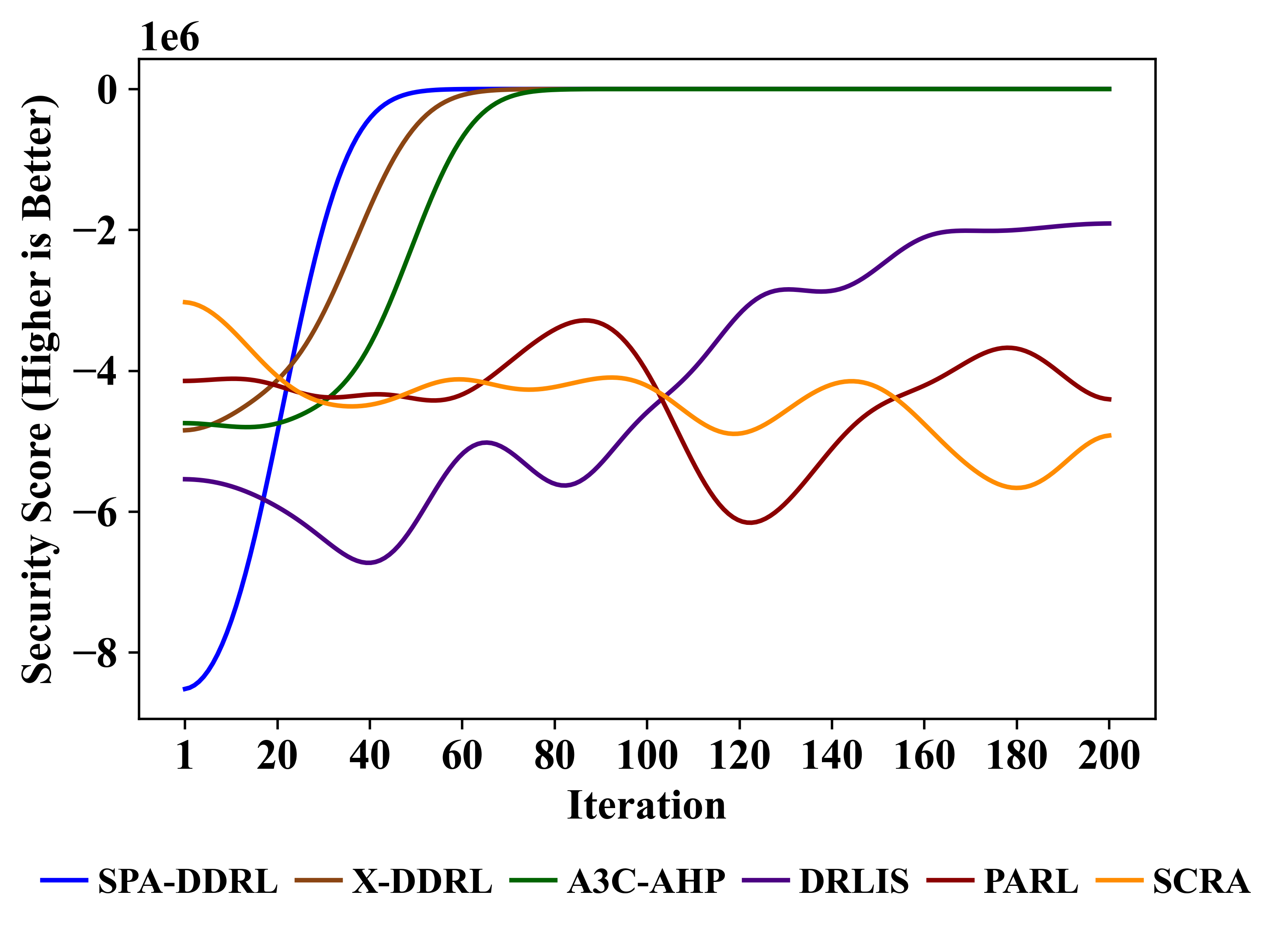}
    \caption{Security Score Convergence}
    \label{fig:security_score_convergence}
\end{subfigure}
\hfill
\begin{subfigure}[b]{0.33\textwidth}
    \centering
    \includegraphics[width=\textwidth]{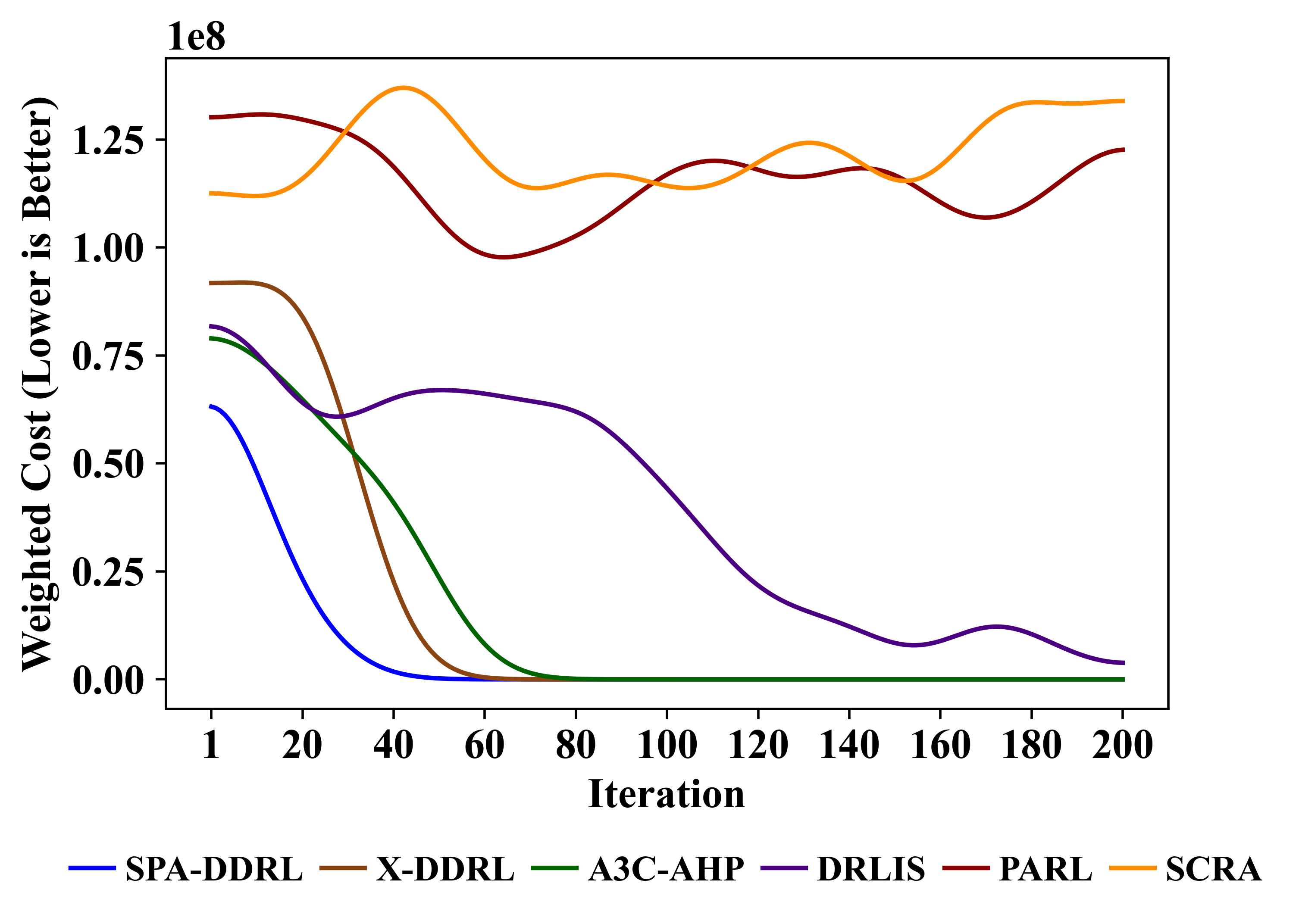}
    \caption{Weighted Cost Convergence}
    \label{fig:weighted_cost_convergence}
\end{subfigure}
\caption{Convergence analysis of SPA-DDRL and baselines for (a) response time, (b) security score, and (c) weighted cost.}
\label{fig:convergence_analysis}
\end{figure*}

\subsubsection{Convergence Analysis}
This section analyzes the convergence behavior and optimization performance of different service placement techniques in terms of response time, security score, and overall weighted cost. The experimental setup involves training on services with task counts $K \in \{5, 10, 20, 80, 100\}$ and evaluating on services containing $K = 40$ tasks, ensuring that the evaluation complexity lies between the training ranges to assess technique robustness on unseen service configurations. Figures~\ref{fig:response_time_convergence}, \ref{fig:security_score_convergence}, and \ref{fig:weighted_cost_convergence} illustrate the convergence trends over 200 iterations for each objective.

\textbf{Response Time:} As shown in Figure~\ref{fig:response_time_convergence}, SPA-DDRL converges to approximately 360 ms, significantly outperforming X-DDRL at 430 ms (16.3\% improvement), A3C-AHP at 950 ms (62.1\% improvement), and DRLIS at 1050 ms (65.7\% improvement). PARL and SCRA fail to converge effectively, exhibiting persistent oscillations and fluctuating around 1300 ms and 1350 ms respectively, representing 72.3\% and 73.3\% performance degradation compared to SPA-DDRL. In terms of convergence speed, SPA-DDRL stabilizes at approximately iteration 40, while X-DDRL requires approximately iteration 60, A3C-AHP requires approximately iteration 70, and DRLIS requires approximately iteration 200 after sustained oscillations.
    
\textbf{Security Score:} Figure~\ref{fig:security_score_convergence} reveals that SPA-DDRL converges to the near-optimal security score at approximately iteration 50, indicating full satisfaction of hard security controls. X-DDRL and A3C-AHP also achieve near-optimal feasible solutions, but require iterations 60 and 70 respectively. In contrast, DRLIS ends at $-2 \times 10^6$, demonstrating fundamental inability to satisfy security requirements. PARL and SCRA exhibit severe security violations, continuously oscillating around $-4 \times 10^6$ and $-6 \times 10^6$ throughout training, indicating their placement strategies persistently violate critical control-level constraints in the three-tier security hierarchy.
    
\textbf{Weighted Cost:} Figure~\ref{fig:weighted_cost_convergence} demonstrates SPA-DDRL's superior joint objective optimization capability, converging to near-zero weighted cost at approximately iteration 50. X-DDRL and A3C-AHP converge to near-zero at iterations 60 and 70 respectively, demonstrating suboptimal but still feasible solutions. DRLIS reduces to $0.05 \times 10^8$ at iteration 200 but still fails to achieve true feasibility due to persistent security violations. PARL and SCRA incur prohibitively high costs, continuously fluctuating around $1.22 \times 10^8$ and $1.32 \times 10^8$ throughout training, with security penalty terms dominating their objective functions.

These results demonstrate that only SPA-DDRL, X-DDRL, and A3C-AHP successfully balance security-performance trade-offs, with SPA-DDRL achieving optimal performance and fastest convergence through synergistic integration of distributed learning, LSTM, and PER.

\subsubsection{System Size Analysis}
\begin{figure*}[!t]
\centering
\begin{subfigure}[b]{\linewidth}
    \centering
    \includegraphics[width=\linewidth, height=3.5cm]{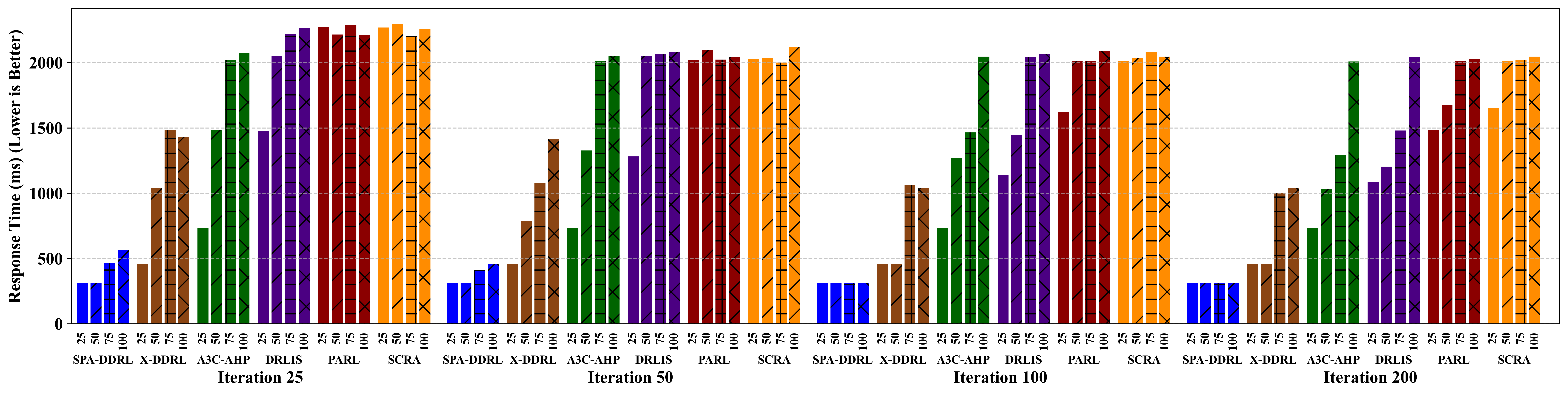}
    \caption{Response Time}
    \label{fig:system_response_time}
\end{subfigure}
\begin{subfigure}[b]{\linewidth}
    \centering
    \includegraphics[width=\linewidth,height=3.5cm]{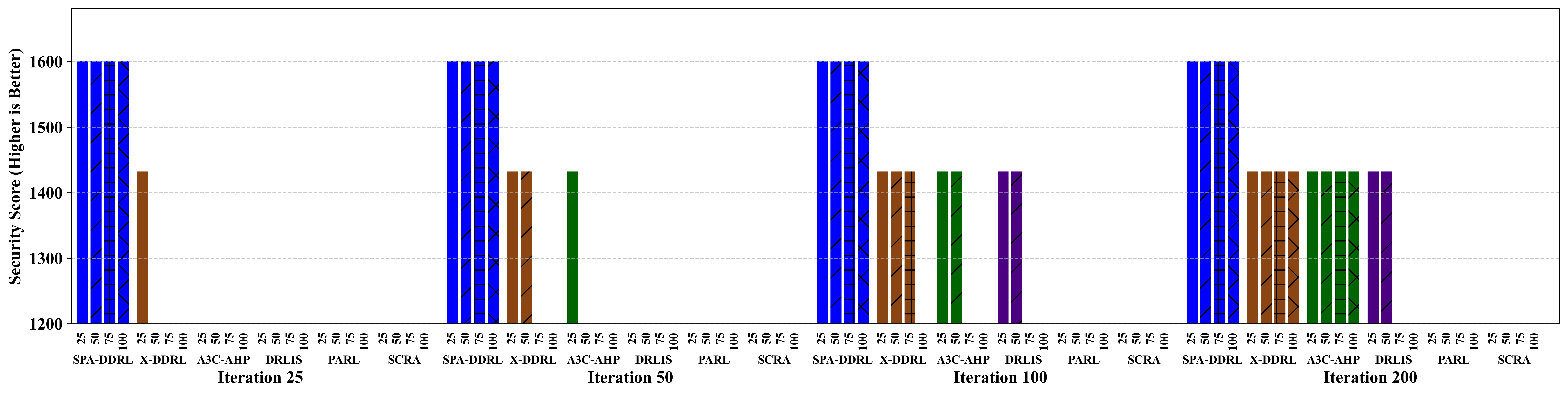}
    \caption{Security Score}
    \label{fig:system_security_score}
\end{subfigure}
\begin{subfigure}[b]{\linewidth}
    \centering
    \includegraphics[width=\linewidth,height=3.5cm]{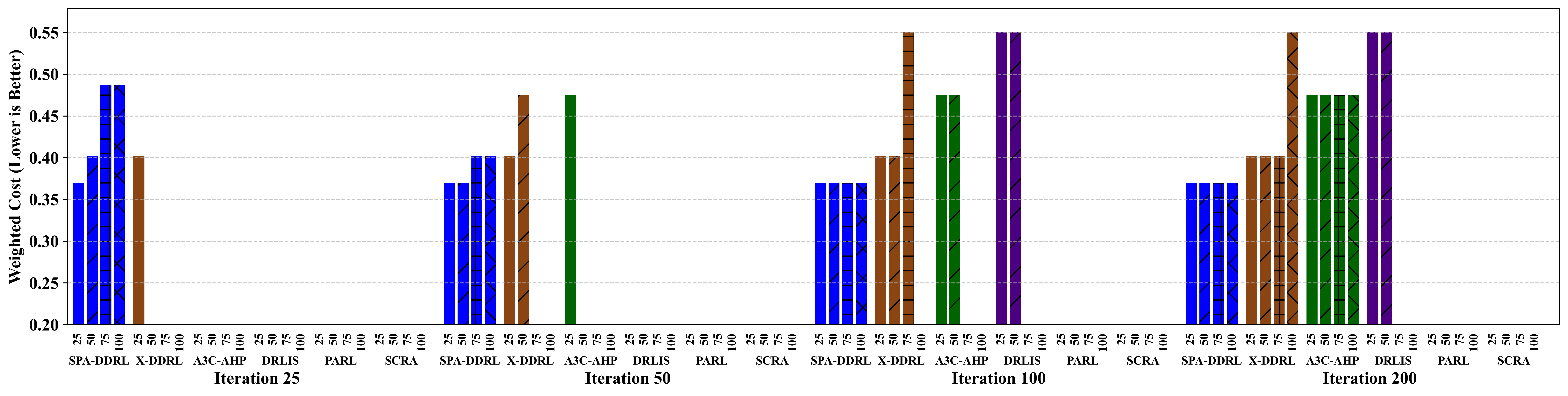}
    \caption{Weighted Cost}
    \label{fig:system_weighted_cost}
\end{subfigure}
\caption{System size analysis comparing techniques performance across different server configurations}
\label{fig:system_size_analysis}
\end{figure*}

This experiment evaluates the performance of service placement techniques when the number of servers increases. The larger number of servers leads to increased search space and complexity of the service placement problem, directly affecting the decision-making process for optimizing both response time and security objectives. In this experiment, we train on services with task counts $K \in \{5, 10, 20, 80, 100\}$ and evaluate on services with $K = 40$ tasks to maintain consistency with previous experiments. In addition, we consider different numbers of servers, where the server count $\in \{25, 50, 75, 100\}$. We evaluate the performance across different iterations $\{25, 50, 100, 200\}$ to observe the convergence behavior under varying system scales. Figures~\ref{fig:system_response_time}, \ref{fig:system_security_score}, and \ref{fig:system_weighted_cost} present the performance results across different system sizes. The absence of bars for certain techniques indicates their failure to find feasible solutions that satisfy the hard security controls under specific system configurations.

\textbf{Response Time:} As shown in Figure~\ref{fig:system_response_time}, SPA-DDRL consistently achieves the lowest response times across all system configurations. As training iterations increase, all techniques show performance improvement. At iteration 200, SPA-DDRL converges to around 300 ms, X-DDRL to 400-1100 ms, A3C-AHP to 700-2000 ms, while DRLIS, PARL, and SCRA exhibit response times in the ranges of 1000-2100 ms, 1400-2100 ms, and 1500-2100 ms respectively, representing 80-85\% performance degradation compared to SPA-DDRL. Within the same iteration, as the number of servers increases from 25 to 100, response times generally rise due to expanded search space and feature complexity. SPA-DDRL demonstrates superior scalability, with response times stabilizing at iterations 100 and 200. In contrast, baseline methods exhibit significant performance degradation with increasing server count: X-DDRL increases from below 500 ms to over 1000 ms, A3C-AHP from 700 ms to 2000 ms, DRLIS from 1000 ms to 2000 ms, PARL from 1500 ms to 2000 ms, and SCRA from 1600 ms to 2000 ms, indicating poor scalability of baseline methods in large-scale systems. 
    
\textbf{Security Score:} As shown in Figure~\ref{fig:system_security_score}, SPA-DDRL uniquely maintains feasible security-compliant solutions across all system configurations and iterations, consistently achieving security scores around 1600. At iteration 25, SPA-DDRL is the only method producing valid solutions across all server counts, while X-DDRL shows limited feasibility for only the 25-server configuration with a lower security score (around 1450). By iteration 50, SPA-DDRL maintains full feasibility, X-DDRL expands to more configurations (25 and 50 servers), and A3C-AHP begins emerging with partial feasibility (25 servers). At iterations 100 and 200, SPA-DDRL continues to demonstrate complete feasibility across all scales, while X-DDRL, A3C-AHP, and DRLIS achieve feasibility for some configurations with security scores around 1450. Throughout all iterations and system scales, PARL and SCRA show a complete absence of feasible solutions, confirming their fundamental inability to satisfy hard security controls in heterogeneous Fog environments. 

\textbf{Weighted Cost:} As shown in Figure~\ref{fig:system_weighted_cost}, SPA-DDRL demonstrates superior scalability in joint security-performance optimization. The feasibility evolution reveals progressive convergence patterns. At iteration 25, SPA-DDRL is the only technique producing feasible solutions across all server counts, with weighted costs ranging from 0.35-0.50. By iteration 50, X-DDRL and A3C-AHP emerge with feasibility for some configurations, achieving costs around 0.40-0.55, while SPA-DDRL optimizes the cost to 0.36-0.40. At iterations 100 and 200, SPA-DDRL stabilizes at weighted costs of 0.36 across all system scales, while X-DDRL ranges from 0.40-0.55, A3C-AHP around 0.47, and DRLIS achieves limited feasibility at 0.55. Throughout all iterations and configurations, PARL and SCRA show a complete absence of feasible solutions, with missing bars indicating persistent failure to satisfy hard security constraints. 

These results demonstrate that SPA-DDRL uniquely maintains solution feasibility and optimal performance across increasing system scales, while baseline techniques progressively show performance decrease and fail to satisfy hard security constraints as environment complexity grows. This superior scalability stems from SPA-DDRL's distributed broker-learner architecture enabling parallel exploration, LSTM-enhanced temporal modeling capturing complex dependencies, and PER facilitating efficient learning from diverse placement scenarios in large-scale heterogeneous Fog computing deployments.

\begin{figure}[htbp]
\centering
\includegraphics[width=1\columnwidth, height=3cm]{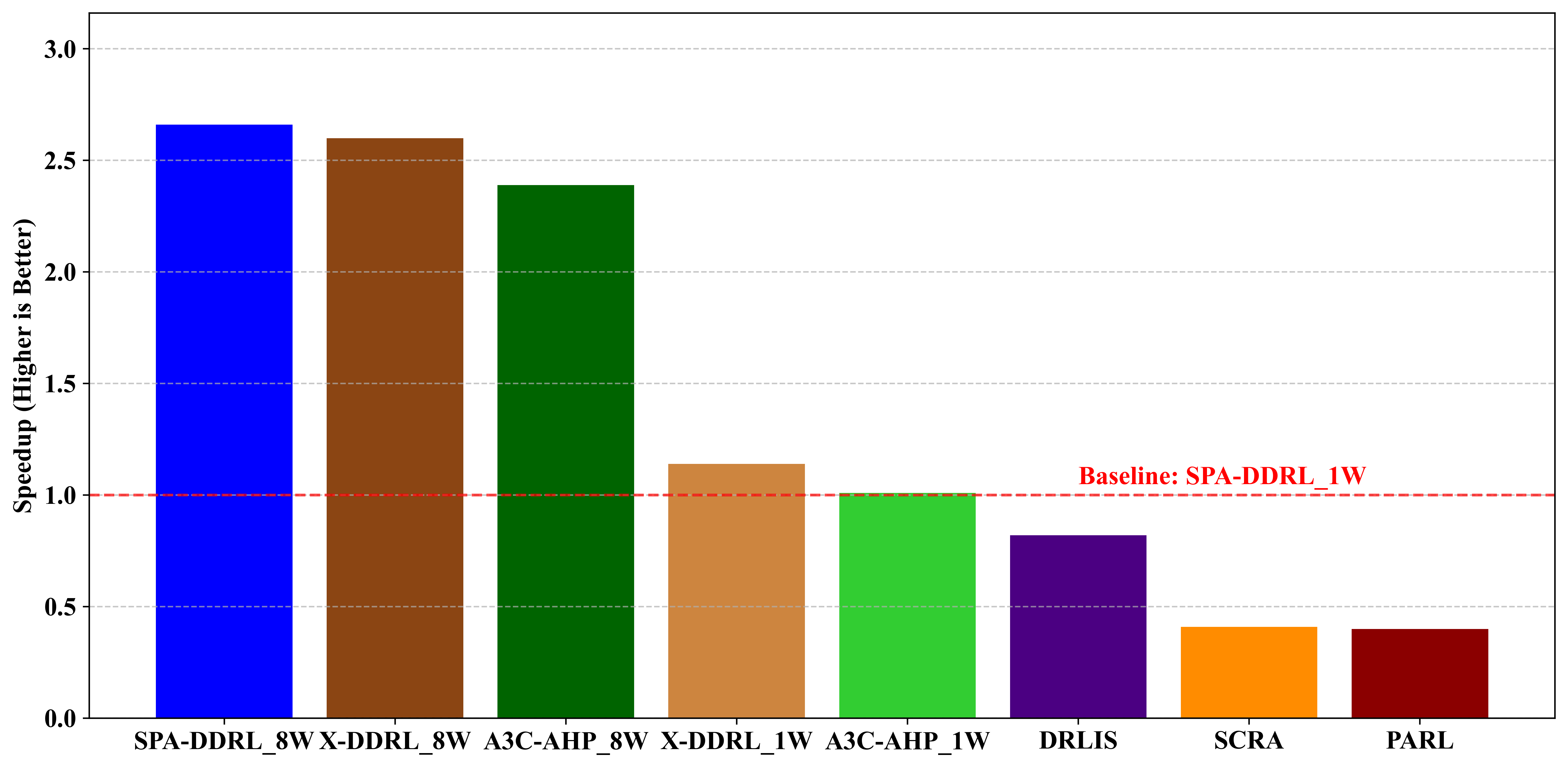}
\caption{Speedup analysis with varying number of workers}
\label{fig:speedup}
\end{figure}

\subsubsection{Speedup Analysis}
This analysis evaluates the efficiency of different techniques in acquiring a predefined number of experience trajectories during training. Faster interactions with the Fog computing environment enable the collection of more diverse experiences, thereby accelerating convergence. The speedup is:
\begin{equation}
    SP=\frac{Time_R}{Time_T}
\end{equation}
where $Time_R$ denotes the time taken by SPA-DDRL with a single worker to reach 150000 environment steps, and $Time_T$ is the time taken by the evaluated technique to reach the same number of steps.

As shown in Figure \ref{fig:speedup}, SPA-DDRL with 8 workers achieves a speedup of approximately 2.7$\times$, outperforming other distributed techniques. X-DDRL and A3C-AHP reach 2.6$\times$ and 2.4$\times$ speedups, respectively, while their single-worker variants exhibit limited improvement. Traditional non-distributed methods such as DRLIS, SCRA, and PARL perform worse than the baseline (speedup $<$ 1.0), indicating slower training progress. The superior speedup of SPA-DDRL is attributed to its effective parallelization, allowing multiple workers to explore distinct action spaces while sharing prioritized experiences. This design can significantly reduce training time and supports rapid adaptation in dynamic Fog computing environments, making SPA-DDRL well-suited for real-world deployments requiring fast convergence and responsiveness.

\subsubsection{Decision Time Overhead Analysis}
This experiment evaluates the average Decision Time Overhead (DTO) for each service placement technique. DTO is defined as the average time required to make a placement decision for each service during the scheduling process.

\begin{figure}[!t]
\centering
\includegraphics[width=1\columnwidth,height=3cm]{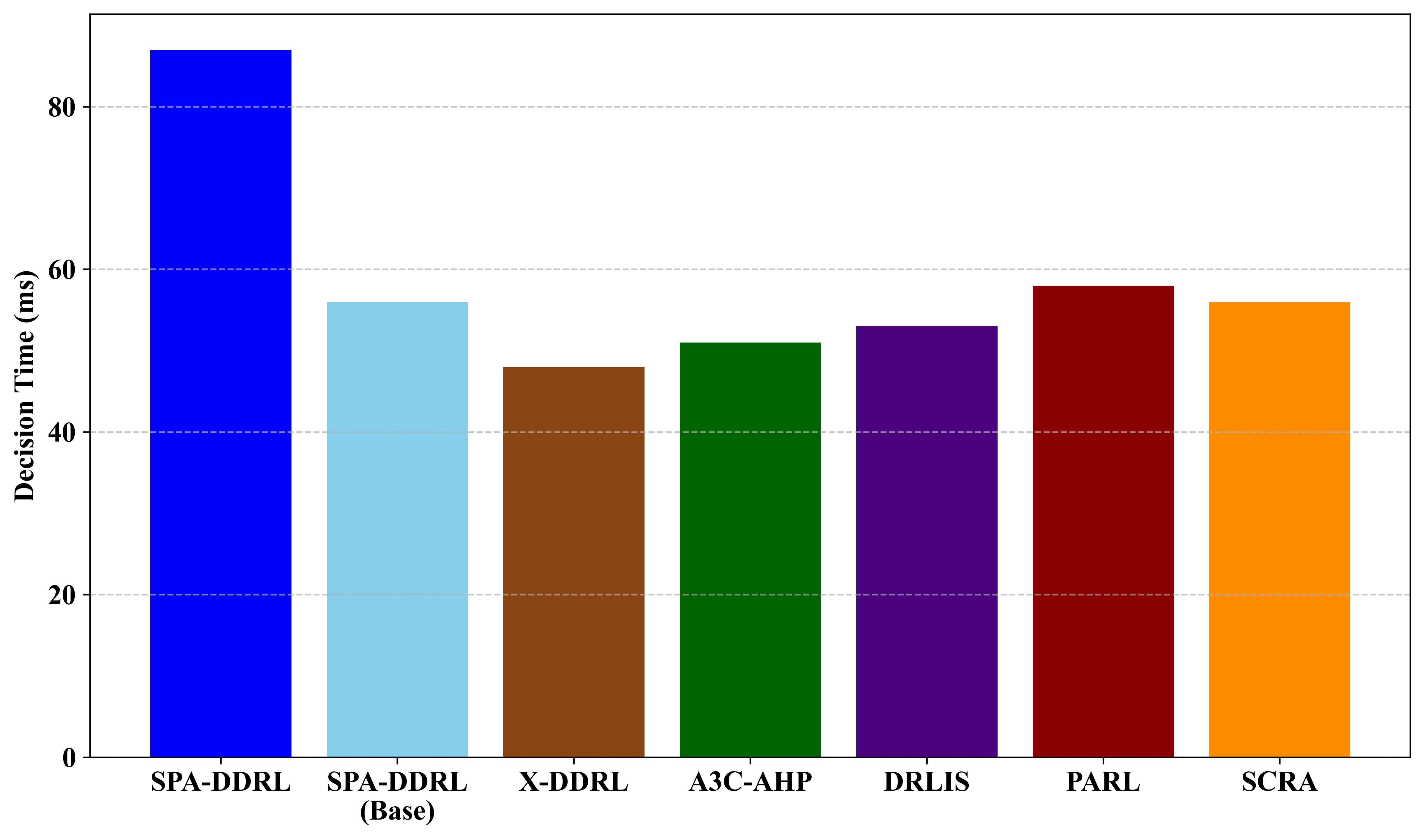}
\caption{Decision Time Overhead analysis}
\label{fig:decision_time}
\end{figure}

As shown in Figure \ref{fig:decision_time}, SPA-DDRL incurs the highest DTO at approximately 87 ms, compared to 56 ms for SPA-DDRL(Base), and 48–58 ms for X-DDRL, A3C-AHP, DRLIS, PARL, and SCRA. The increased overhead in SPA-DDRL is mainly due to the computational cost introduced by the LSTM for temporal dependency modeling and the PER, which involves analyzing historical states and ranking experiences for sampling. Although SPA-DDRL has a higher DTO, the overhead is acceptable given its superior convergence speed, solution quality, and scalability, as demonstrated in prior experiments. The modest increase in DTO is well justified by the substantial performance gains it delivers in optimization and scalability performance.

\section{Conclusions and Future Work}
\label{conclusion}
This paper presented SPA-DDRL, a distributed framework designed to jointly optimize security compliance and response time within heterogeneous Fog computing environments. We formulated a dual-objective optimization model underpinned by a novel three-tier security quantification scheme—comprising configuration, capability, and control-level assessments—to rigorously enforce stringent policies such as data encryption and access control. By embedding this quantification into the reward function, the framework ensures compliant service placement even under dynamic conditions. Furthermore, the proposed distributed broker-learner architecture, augmented by LSTM networks for temporal modeling, Prioritized Experience Replay (PER), and off-policy correction, effectively reconciles training stability with scalability. Extensive evaluations demonstrate the superiority of SPA-DDRL over state-of-the-art baselines in terms of convergence speed and solution quality. Crucially, the framework maintains solution feasibility and strict security compliance as system scale increases, addressing a critical failure mode of competing methods that often yield non-compliant solutions. These capabilities validate SPA-DDRL as a robust solution for mission-critical, secure deployments in large-scale Fog ecosystems.

As part of future work, we plan to extend SPA-DDRL to handle federated learning scenarios where privacy preservation is critical alongside security and performance objectives. Moreover, we intend to explore adaptive security requirement adjustment mechanisms that can dynamically modify security levels based on real-time threat intelligence and evolving attack patterns in Fog computing environments, further enhancing its security resilience.



\bibliographystyle{IEEEtran}
\bibliography{ref}

\end{document}